\documentclass[fleqn,usenatbib]{mnras}

\usepackage{amsfonts}

\usepackage[T1]{fontenc}
\usepackage{ae,aecompl}
\usepackage[position=top]{subfig}
\usepackage{float}
\usepackage{comment}
\usepackage[normalem]{ulem}

\usepackage{booktabs}
\usepackage[position=top]{subfig}
\usepackage{grffile}
\usepackage{etoolbox}
\makeatletter 
  \patchcmd{\NAT@citex}
    {\@citea\NAT@hyper@{%
      \NAT@nmfmt{\NAT@nm}%
      \hyper@natlinkbreak{\NAT@aysep\NAT@spacechar}{\@citeb\@extra@b@citeb}%
      \NAT@date}}
    {\@citea\NAT@nmfmt{\NAT@nm}%
    \NAT@aysep\NAT@spacechar\NAT@hyper@{\NAT@date}}{}{}

  \patchcmd{\NAT@citex}
    {\@citea\NAT@hyper@{%
      \NAT@nmfmt{\NAT@nm}%
      \hyper@natlinkbreak{\NAT@spacechar\NAT@@open\if*#1*\else#1\NAT@spacechar\fi}%
        {\@citeb\@extra@b@citeb}%
      \NAT@date}}
    {\@citea\NAT@nmfmt{\NAT@nm}%
    \NAT@spacechar\NAT@@open\if*#1*\else#1\NAT@spacechar\fi\NAT@hyper@{\NAT@date}}
    {}{}
\makeatother

\defcitealias{Brown2019}{B19}

 \makeatother
\usepackage{graphicx}	
\usepackage{amsmath}	
\usepackage{amssymb}	
\usepackage{newtxtext,newtxmath}
\usepackage{pifont}
\usepackage{xcolor}
\hypersetup{allcolors=orange}

\title[Analytical model for Pop~III star formation]{Towards a universal analytical model for Population~III star formation: interplay between feedback and fragmentation}

\author[B. Liu et al.]{Boyuan Liu\textsuperscript{\href{https://orcid.org/0000-0002-4966-7450}{\includegraphics[width=2.5mm]{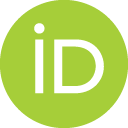}}\,}\thanks{E-mail: treibeis1995@gmail.com}$^{1,2}$, James Gurian\textsuperscript{\href{https://orcid.org/0000-0002-8677-1038}{\includegraphics[width=2.5mm]{orcid.png}}\,}$^{3}$, Kohei Inayoshi\textsuperscript{\href{https://orcid.org/0000-0001-9840-4959}{\includegraphics[width=2.5mm]{orcid.png}}\,}$^{4}$, Shingo Hirano\textsuperscript{\href{https://orcid.org/0000-0002-4317-767X}{\includegraphics[width=2.5mm]{orcid.png}}\,}$^{5}$, Takashi Hosokawa\textsuperscript{\href{https://orcid.org/0000-0003-3127-5982}{\includegraphics[width=2.5mm]{orcid.png}}\,}$^{6}$, 
\newauthor 
Volker Bromm\textsuperscript{\href{https://orcid.org/0000-0003-0212-2979}{\includegraphics[width=2.5mm]{orcid.png}}\,}$^{7,8}$, 
and Naoki Yoshida\textsuperscript{\href{https://orcid.org/0000-0001-7925-238X}{\includegraphics[width=2.5mm]{orcid.png}}\,}$^{9,10,11}$\\
$^{1}$Institute of Astronomy, University of Cambridge, Madingley Road, Cambridge, CB3 0HA, UK\\
$^{2}$Institut für Theoretische Astrophysik, Zentrum für Astronomie, Universität Heidelberg, Albert Ueberle Straße 2, D-69120 Heidelberg, Germany\\
$^{3}$Perimeter Institute for Theoretical Physics, Waterloo, Ontario, N2L 2Y5, Canada\\
$^{4}$Kavli Institute for Astronomy and Astrophysics, Peking University, Beijing 100871, People’s Republic of China\\
$^{5}$Department of Astronomy, School of Science, University of Tokyo, Tokyo 113-0033, Japan\\
$^{6}$Department of Physics, Kyoto University, Sakyo, Kyoto 606-8502, Japan\\
$^{7}$Department of Astronomy, University of Texas, Austin, TX 78712, USA\\
$^{8}$Weinberg Institute for Theoretical Physics, University of Texas, Austin, TX 78712, USA\\
$^{9}$Department of Physics, School of Science, The University of Tokyo, 7-3-1 Hongo, Bunkyo, Tokyo 113-0033, Japan\\
$^{10}$Research Center for the Early Universe, School of Science, The University of Tokyo, 7-3-1 Hongo, Bunkyo, Tokyo 113-0033, Japan\\
$^{11}$Kavli Institute for the Physics and Mathematics of the Universe (WPI), The University of Tokyo, Kashiwa, Chiba 277-8583, Japan
}

\date{Accepted XXX. Received YYY; in original form ZZZ}

\pubyear{2024}

\begin{document}
\label{firstpage}
\pagerange{\pageref{firstpage}--\pageref{lastpage}}
\maketitle

\begin{abstract}
JWST has brought us new insights into Cosmic Dawn with tentative detection of the unique signatures of metal-free Population~III (Pop~III) stars, such as strong HeII emission, extremely blue UV spectrum, and enhanced nitrogen abundance. 
Self-consistent theoretical predictions of the formation rates, sites, and masses of Pop~III stars are crucial for interpreting the observations, but are challenging due to complex physical processes operating over the large range of length scales involved. One solution is to combine analytical models for the small-scale star formation process with cosmological simulations that capture the large-scale physics such as structure formation, radiation backgrounds, and baryon-dark matter streaming motion that regulate the conditions of Pop~III star formation. We build an analytical model to predict the final masses of Pop~III stars/clusters from the properties of star-forming clouds, based on the key results of small-scale star formation simulations and stellar evolution models. Our model for the first time considers the interplay between feedback and fragmentation and covers different modes of Pop~III star formation ranging from ordinary small ($\sim 10-2000\ \rm M_\odot$) clusters in molecular-cooling clouds to massive ($\gtrsim 10^{4}\ \rm M_\odot$) clusters containing supermassive ($\sim 10^{4}-3\times 10^{5}\ \rm M_\odot$) 
stars under violent collapse of atomic-cooling clouds with large gas accretion rates of 
$\gtrsim 0.1\ \rm M_\odot\ yr^{-1}$. As an example, the model is applied to the Pop~III star-forming clouds in the progenitors of typical haloes hosting high-$z$ luminous quasars ($M_{\rm h}\sim 10^{12}\ \rm M_\odot$ at $z\sim 6$), which shows that formation of Pop~III massive clusters is common ($\sim 20-70\%$) in such biased ($\sim4\sigma$) regions, and the resulting heavy black hole seeds from supermassive stars can account for a significant fraction of observed luminous ($\gtrsim 10^{46}\ \rm erg\ s^{-1}$) quasars at $z\sim 6$. 
\end{abstract}



\section{Introduction}
The first generation of so-called Population~III (Pop~III) stars, formed in metal-free primordial gas, are expected to be fundamentally different from present-day metal-enriched Population~I/II (Pop~I/II) stars \citep[reviewed by, e.g.,][]{Bromm2009,Bromm2013,loeb2013,Haemmerle2020,Klessen2023}. Due to their pristine formation environments, Pop~III stars tend to form in small clusters with a broad top-heavy initial mass function (IMF) extending up to a few $\sim 10^{5}\ \rm M_{\odot}$ according to small-scale (radiative) (magneto-) hydrodynamic simulations \citep[e.g.][]{Clark2011,Greif2011,Greif2012,Susa2014,Hirano2014,Hirano2015,Hirano2018sc,Stacy2013,Stacy2014,Stacy2016,Hosokawa2016,Hirano2017,Hirano2017sms,Suazo2019,Susa2019,Sugimura2020,Sugimura2023,McKee2020,Wollenberg2020,Park2021,Park2022,Park2024,Chon2018,Chon2021,Chon2022,Chon2020,Sakurai2020,Regan2014,Regan2017,Regan2020,Sharda2020,Sharda2021,Latif2021,Latif2022,Woods2021,Woods2024,Riaz2018,Riaz2023,Riaz2022sf,Prole2022res,Prole2022,Prole2023,Prole2023bh,Regan2023,Reinoso2023,Toyouchi2023,Kiyuna2024,Sharda2024,Sadanari2024}. Further, stellar evolution models predict that Pop~III stars are more compact and hotter than Pop~I/II stars, likely fast rotating, have negligible mass loss from metal line-driven winds \citep[e.g.][]{Schaerer2002,Meynet2006,Ekstrom2008,Heger2010,Yoon2012,Tanikawa2020,Farrell2021,Murphy2021,Aryan2023,Martinet2023,Nandal2023,Volpato2023}, and end their lives in diverse supernova events with distinct metal yields or collapse directly into massive back holes (BHs) \citep[e.g.,][]{Heger2002,Heger2003,Maeda2003,Umeda2003,Umeda2005,Iwamoto2005,Kobayashi2006,Tominaga2009,Heger2010,Nomoto2013}. 

With these peculiar properties, Pop~III stars play important roles in the first billion years of cosmic history (i.e., Cosmic Dawn) through their radiation, metal enrichment, and by seeding BHs, which produce unique signatures in direct observations \citep[e.g.,][]{Windhorst2018,Grisdale2021,Nakajima2022,Vikaeus2022,Trussler2023,Katz2023,Larkin2023,Venditti2024,Zackrisson2011,Zackrisson2012,Zackrisson2023}, imprints in cosmic chemical and thermal evolution (see, e.g., \citealt{Karlsson2013,Nomoto2013,Frebel2015,Barkana2016,Dayal2018}, for reviews), and impact on the formation and evolution of subsequent populations of stars, galaxies and supermassive BHs (SMBHs) 
\citep[see, e.g.,][]{Bromm2011,Johnson2013,Pawlik2013,Jeon2015,Smith2015,Regan2017,Sakurai2017,Haemmerle2020,Inayoshi2020,Schauer2021,Volonteri2021,Chon2022,Sarmento2022,Chiaki2023,Sanati2023,Reinoso2023,Wise2023,Kiyuna2024,Regan2024,Rossi2024}, 
allowing us to constrain their properties through direct and indirect observations. 

Recently, JWST has revealed the Cosmic Dawn with unprecedented discoveries of puzzling phenomena unexpected from conventional theoretical models \citep{Adamo2024}, some of which can be explained by the unique signatures of Pop~III stars and their BH remnants, such as metal-poor strong HeII-emission systems \citep{Wang2022jwst,Vanzella2023,Maiolino2023}, galaxies with extremely blue UV spectra \citep[e.g.,][]{Nanayakkara2023,Austin2024,Cullen2024}, strong carbon and nitrogen enrichment \citep{D'Eugenio2023,Cameron2023,Senchyna2023,Ji2024,Schaerer2024,Topping2024} likely from fast-rotating or very massive Pop~III stars \citep{Liu2021wind,Nandal2024rot,Nandal2024vms,Nandal2024,Tsiatsiou2024}, highly magnified individual massive stars \citep{Welch2022,Schauer2022,Zackrisson2023}, {a slowly evolving UV luminosity function hinting at high UV variability, non-constant star formation efficiency (SFE), extreme nebular emission powered by hot ionizing sources, and top-heavy IMF \citep[e.g.,][]{Inayoshi2022jwst,Finkelstein2023,Shen2023,Conselice2024,Cueto2024,Harikane2024,Katz2024,Trinca2024,Ventura2024}}, and overabundant luminous quasars with overmassive (rapidly spinning) BHs at $z\gtrsim 5$ \citep[e.g.,][]{Goulding2023,Harikane2023,Maiolino2023agn,Maiolino2024,Kokorev2023,Ubler2023,Akins2024,Durodola2024,Greene2024,Juodzbalis2024,Kocevski2024,Matthee2024,Inayoshi2024} likely originating from heavy BH seeds ($\gtrsim 10^{4-5}\ \rm M_\odot$). 
Although there is still no conclusive direct detection of Pop~III-dominated galaxies, which is likely challenging for JWST \citep{Gardner2006,Riaz2022,Nakajima2022,Katz2023,Bovill2024} and may require large extraterrestrial telescopes \citep{Angel2008,Rhodes2020,Schauer2020}, these observations highlight the importance of the Pop~III components in high-$z$ galaxies. As JWST continues its revolutionary observations, other facilities (e.g., Roman Space Telescope and Vera C. Rubin Observatory) will come online in the near future, which may bring us new detection of Pop~III signatures, such as the transient signals of Pop~III gamma-ray bursts and supernova explosions \citep[e.g.,][]{Fryer2022,Lazar2022,Hartwig2023,Venditti2024sn,Wiggins2024}.

Gravitational wave observations of massive binary BH (BBH) mergers have also tentatively discovered a non-negligible contribution from Pop~III BH remnants \citep[up to $10\%$ of the detected events, see, e.g.,][]{Kinugawa2020,Kinugawa2021,Iwaya2023}, especially for the most massive mergers such as GW190521 with unusual BH masses \citep{Abbott2020,abbott2020gw190521} in the range $\sim50-130\ \rm M_{\odot}$ mostly forbidden for BHs from Pop~I/II stars by standard pair-instability supernova (PISN) models \citep[e.g.,][]{Heger2003,Belczynski2016,Yoshida2016,Woosley2017,Spera2017,Marchant2019,Marchant2020,Mapelli2020rotation}. 
Thanks to their massive, compact nature and lack of mass loss, Pop~III stars can produce BHs up to $\sim 100\ \rm M_\odot$ before triggering PISNe and more massive ($\gtrsim 130\ \rm M_\odot$) BHs via direct collapse \citep[e.g.,][]{Farrell2021,Volpato2023,Santoliquido2023,Mestichelli2024}, which can end up in BBH mergers like GW190521 via isolated binary stellar evolution under proper conditions \citep{Kinugawa2021gw,Tanikawa2021,Tanikawa2022,Santoliquido2023,Tanikawa2024} or through dynamical interactions in massive ($\gtrsim 10^4\ \rm M_\odot$) dense clusters of Pop~III stars/BHs themselves \citep{Wang2022,Liu2023sc,Mestichelli2024} and (Pop~I/II) nuclear star clusters \citep{Liu2020gw,Liu2021,Liu2024}. In the next decades, the 3rd-generation GW detectors such as the Einstein Telescope \citep[ET,][]{Punturo2010,Maggiore2020} and the Cosmic Explorer \citep{Reitze2019,Evans2023} will be able to detect BBH mergers up to $z\sim 30$, among which the contribution of Pop~III BH remnants are expected to be larger and better characterized \citep[e.g.,][]{Tanikawa2022,Franciolini2024,Santoliquido2024}.


To fully understand the roles played by Pop~III stars at Cosmic Dawn and the underlying physics (of early star formation, stellar feedback, BH formation and growth) through observations, self-consistent theoretical predictions are required not only for 
\begin{itemize}
    \item[(1)] the formation rates and sites of Pop~III stars, but also for 
    \item[(2)] the masses of Pop~III star clusters and the stellar IMF.
\end{itemize}
Evidently, the total mass of Pop~III stars in a galaxy is vital for the detectability of Pop~III signatures embedded in the signals from co-existing Pop~I/II stars and AGN, which are common in the galaxies detectable by JWST \citep[e.g.,][]{Sarmento2018,Liu2020,Riaz2022,Venditti2023,Venditti2024}. The cluster mass also determines whether dynamical interactions can efficiently produce (distinguishable) massive, eccentric Pop~III BBH mergers \citep{Wang2022,Liu2023sc,Mestichelli2024}, which may not be possible for isolated binary stellar evolution \citep{Costa2023,Santoliquido2023}. The Pop~III IMF affects the UV spectra of Pop~III clusters \citep[e.g.,][]{Zackrisson2011,Zackrisson2012,Bovill2024} and the X-ray outputs of Pop~III X-ray binaries \citep{Sartorio2023}, shaping the imprints of Pop~III stars on the 21-cm signal and reionization {\citep[e.g.,][]{Fialkov2014,Gessey-Jones2022,Gessey-Jones2024,Salvador-Sole2017,Salvador-Sole2022,Fialkov2023}}. 
Moreover, the shape of the Pop~III IMF in the high-mass regime ($\gtrsim 100\ \rm M_\odot$) has essential implications on the rates/abundances of strongly lensed massive Pop~III stars detectable by JWST \citep{Zackrisson2023}, Pop~III PISNe \citep[e.g.,][]{Lazar2022,Wiggins2024,Venditti2024sn}, very massive stars (VMSs, $\sim 10^3-10^4\ \rm M_\odot$) with peculiar metal enrichment \citep[e.g.,][]{Nandal2024vms,Nandal2024}, and supermassive stars (SMSs, $\gtrsim 10^{4}\ \rm M_\odot$) as progenitors of heavy BH seeds \citep{Smith2019,Inayoshi2020,Haemmerle2020,Volonteri2021,Wise2023,Regan2024} that are likely needed to explain the SMBHs in high-$z$ luminous quasars \citep[e.g.,][]{Fan2001,Willott2010,Matsuoka2016,Onoue2019,Shen2020,Yang2021,Fan2023,Akins2024,Durodola2024,Greene2024,Matthee2024,Kocevski2024}. 

These two aspects are expected to be closely related to each other in the rapidly evolving environments of Cosmic Dawn. However, they are usually modelled separately in current theoretical studies, as it is challenging to combine them in one theoretical framework due to the large range of scales involved (from over-Mpc to sub-AU). The first aspect is mostly investigated with large-scale cosmological hydrodynamic simulations \citep[e.g.,][]{Johnson2013,Smith2015,Xu2016,Sarmento2017,Sarmento2018,El-Badry2018,Liu2019,Liu2020sim,Liu2020,Liu2022,Skinner2020,Kulkarni2021,Schauer2019vbc,Schauer2021,Kulkarni2022,Yajima2022,Yajima2023,Kiyuna2023,Garcia2023,Venditti2023,Incatasciato2023,CorreaMagnus2024,Lenoble2024,Sugimura2024,Smith2024} and semi-analytical models \citep[e.g.,][]{Manrique2015,Salvador-Sole2017,Griffen2018,Dayal2020,Visbal2020,Li2021,Lupi2021,Hartwig2022,Trinca2022,Trinca2024,Hegde2023,Nebrin2023,Bovill2024,Ventura2024,Feathers2024}, which show that the formation rates and sites of Pop~III star formation are regulated by cosmic/halo-scale processes, such as cosmic structure formation (for different dark matter models), dissociation of $\rm H_2$ by Lyman-Werner (LW) radiation, streaming motion between baryons and dark matter, ionization and heating by X-rays, dynamical heating by halo mergers, and metal enrichment. For instance, LW radiation, streaming motion, and halo mergers can delay Pop~III star formation, even shifting the formation sites to more massive ($\gtrsim 10^8\ \rm M_\odot$) atomic-cooling haloes \citep[e.g.][]{Yoshida2003,Tseliakhovich2010,Tseliakhovich2011,Fialkov2012,Tanaka2014,Fernandez2014,Schauer2017,Inayoshi2018,Regan2020cos,Kulkarni2021,Schauer2021}. Such large-scale models lack the resolution to follow the sub-pc scale star formation process in detail (which is computationally prohibitive in a cosmologically representative volume). As a result, they estimate the final outcomes of Pop~III star-formation with simple assumptions on the SFE and IMF treated as independent global parameters. On the other hand, for the second aspect, small-scale high-resolution (idealized/zoom-in) simulations of star-forming clouds (see above for references) have revealed 
that the final mass of Pop~III stars formed and their IMF are determined by the complex interplay between fragmentation, radiative feedback, and stellar evolution, which have non-trivial correlations with the (initial) properties of the star-forming cloud. For instance, the standard pathway of Pop~III star formation in molecular-cooling minihaloes ($M_{\rm h}\sim 10^{5-6}\ \rm M_\odot $ at $z\sim 20-30$ with gas infall rates $\sim 10^{-4}-0.01\ \rm M_\odot\ yr^{-1}$) typically produces Pop~III stars with masses $\sim 10-1000\ \rm M_\odot$ regulated by radiative feedback \citep[e.g.,][]{Hirano2014,Hirano2015,Park2021,Sugimura2023}, while more violent collapse ($\gtrsim 0.1\ \rm M_\odot\ yr^{-1}$) in atomic-cooling haloes can overcome/suppress radiative feedback and produce SMSs up to a few $10^{5}\ \rm M_\odot$ \citep[e.g.,][]{Bromm2003,Chon2018,Chon2020,Sakurai2020,Toyouchi2023,Reinoso2023,Regan2023,Kiyuna2024}.

To bridge the aforementioned insights from theoretical efforts on the (1) large- and (2) small-scale aspects and achieve self-consistent modelling of Pop~III star formation across all scales, we build a universal analytical model to predict the outcomes of Pop~III star formation from the properties of star-forming clouds, as described in Sec.~\ref{sec:method}. Our model inherits the basic elements from previous analytical studies \citep[e.g.,][]{Omukai2002,McKee2002,McKee2003,McKee2008,Tan2004,Hosokawa2009,Johnson2012,Yajima2017,Fukushima2018,Li2021,Toyouchi2023} such as the balance between mass inflow from the collapsing cloud and mass loss by feedback, the reaction of protostars to accretion, 
as well as their finite lifetimes and maximum masses set by stellar evolution (Sec.~\ref{sec:mass}). It captures the key trends in the evolution of Pop~III protostar systems inferred from small-scale simulations with simple scaling relations governed by physically motivated parameters (Sec.~\ref{sec:scale}). Furthermore, our model can track the growth and feedback of multiple protostars, and is used to explore the interplay between feedback and fragmentation in determining the formation efficiency and IMF of Pop~III stars (Sec.\ref{sec:multiplicity}, Appendices~\ref{apdx:details} and \ref{apdx:param}). 
As an example, the model is applied to the Pop~III star-forming clouds in the progenitor haloes of high-$z$ luminous quasar host galaxies \citep{Li2021} to evaluate the mass distributions of Pop~III stars/clusters formed in such over-dense regions (Sec.~\ref{sec:mdis}). We also discuss the caveats in our model and possible ways to improve it (Sec.~\ref{sec:caveats}). Given its simplicity and flexibility, our model can be easily incorporated into large-scale semi-analytical models and cosmological simulations.

\section{Analytical model}
\label{sec:method}
In this section, we describe our analytical model for Pop~III star formation. We first introduce the scaling relations learned from simulations to capture the general evolution of Pop~III star-forming discs and protostar systems governed by gravity and hydrodynamics (Sec.~\ref{sec:scale}). Next, we consider the roles played by feedback and stellar evolution, which are combined with the scaling relations to calculate the final mass under the simple assumption that only a single star forms per cloud (Sec.~\ref{sec:mass}). Finally, we generalize the model to consider multiple protostars and explore the effects of fragmentation/multiplicity on the final stellar mass (Sec.~\ref{sec:multiplicity}). For conciseness, we only show the results for select examples below. To demonstrate the flexibility of our model, we provide more detailed results for a broader exploration of parameter space and an alternative parametrisation of multiplicity in Appendices~\ref{apdx:details} and \ref{apdx:param}. The qualitative conclusions shown in this section remain valid for other choices of parameters/assumptions, although the predicted final stellar mass (for the same gas infall rate) can vary up to an order of magnitude. 

\subsection{Scaling relations for Pop~III star-forming discs and protostar systems}
\label{sec:scale}
It is shown in \citet[see their sec.~2.1]{Liu2021binary} that the total masses $M$ and radii $R$ of Pop~III protostar systems in hydrodynamic simulations of typical primordial star-forming clouds in minihaloes ($M_{\rm h}\sim 10^{5}-10^{6}\ \rm M_\odot$, $z\sim 20-30$) can be reproduced with an analytical solution within a factor of $3$ at least in the early stage \citep[see also fig.~9 in][]{Susa2019}: 
\begin{align}
    M&\simeq 400\ \mathrm{M_{\odot}}\ \left[t/\left(10^{5}\ \rm yr\right)\right]^{\beta}\ ,\label{m_t}\\
    R&\simeq \mathrm{AU}\ (t/ \mathrm{yr})^{\delta}\ ,\label{r_t}
\end{align}
where $\beta=4-3\gamma_{\rm eff}=0.73$, $\delta=2-\gamma_{\rm eff}=0.91$, given $\gamma_{\rm eff}=1.09$ as the polytropic index in the effective equation of state (EoS) $P\propto n^{\gamma_{\rm eff}}$ of collapsing primordial gas under canonical $\rm H_2$ cooling\footnote{The EoS can be different in non-standard pathways of Pop~III star formation on lower- or higher-temperature tracks of primordial gas collapse driven by HD or atomic-cooling \citep[e.g.,][]{Hirano2014,Regan2023,Gurian2024}. For simplicity, we adopt the standard value $\gamma_{\rm eff}=1.09$ for $\rm H_2$ cooling by default and explore the dependence of our results on $\gamma_{\rm eff}$ in Appendix~\ref{apdx:scal}.} in the temperature range $T\sim 300-1000\ \rm K$ \citep{Omukai1998,Omukai2005}. To generalize this solution for Pop~III star formation in smaller/larger haloes with weaker/stronger gas inflows, we re-write Eq.~\ref{m_t} with a normalisation factor $M_{0}$ and a characteristic timescale $t_{0}$, and express $R$ with $M$:
\begin{align}
    M&=M_{0}(t/t_{0})^{\beta}\ ,\label{m_t_new}\\
    R&=\left[10^{5\beta}M/(400\ {\rm M_\odot})\right]^{\delta/\beta}\ {\rm AU}\propto M^{\delta/\beta}\simeq M^{1.25}\ . \label{r_m_orig}
\end{align}
Taking time derivative of Eq.~\ref{m_t_new}, the accretion rate $\dot{M}$ can be expressed as a function of mass 
$M$: 
\begin{align}
    \dot{M}(M)=(\beta M/t_{0})(M/M_{0})^{-1/\beta}\propto M^{1-1/\beta}\simeq M^{-0.37}\ .\label{mdot_m}
\end{align}
This scaling is also predicted by (spherically symmetric) similarity solutions for polytropic gas with $P\propto n^{\gamma_{\rm eff}}$ given $\gamma_{\rm eff}=1.09$ \citep[e.g.,][]{Yahil1983,Suto1988,Omukai1998,Tan2004}, as well as cosmological zoom-in simulations of the accretion-driven growth of a Pop~III protostar \citep[][]{BrommLoeb2004}.

Next, we set the characteristic timescale $t_0$ using the relation between time and density normalisation parameters in self-similar transformations of the hydrodynamic equations \citep[see eq.~1 in][]{Liu2021binary}
\begin{align}
    t_{0}=1/\sqrt{4\pi G\rho_{0}}\simeq 3\times 10^{4}\ \rm yr\ ,
\end{align}
where $\rho_{0}=\mu m_{\rm H}n_{0}$, given the proton mass $m_{\rm H}$, the mean molecular weight of primordial gas $\mu\simeq 1.22$, and the typical density $n_{0}=10^{6}\ \rm cm^{-3}$ around which primordial gas in run-away collapse under $\rm H_2$ cooling settles to the effective EoS $P\propto n^{1.09}$. Here, $n_{0}=10^{6}\ \rm cm^{-3}$ is also the typical density at the edge of a Pop~III star-forming disc \citep[see, e.g., fig.~8 in][]{Toyouchi2023}. Finally, to determine $M_{0}$, we equate the accretion rate of the protostar at $t_{0}$, $\dot{M}|_{t=t_{0}}=\beta M_{0}/t_{0}$, to the cloud-scale initial gas inflow rate $\dot{M}_{\rm in}$ onto the disc, multiplied by a factor $\eta\sim 0.4-0.6$ \citep{Sakurai2016,Toyouchi2023} that captures the rotational support against collapse, {which gives
\begin{align}
    &M_{0}=\eta\dot{M}_{\rm in}t_{0}/\beta\ ,\label{m_0}\\
    &M(t)=(\eta/\beta)\dot{M}_{\rm in} t_0^{1-\beta}t^\beta\propto (\eta/\beta)\dot{M}_{\rm in} n_0^{(\beta-1)/2}t^\beta\ .\label{m_t_final}
\end{align}}
The effects of other aspects, such as ejections of protostars by dynamical interactions and magnetic/radiative outflows, can also be absorbed into the parameter $\eta$. Thanks to the simple power-law scaling, once the slope $\beta=4-3\gamma_{\rm eff}$ is known, the mass evolution is completely governed by one parameter, {$A\equiv (\eta/\beta)\dot{M}_{\rm in} t_0^{1-\beta}$}. 
This also indicates that our results are not very sensitive to the disc edge density $n_0$ given the weak dependence of $A\propto t_0^{1-\beta}\propto n_0^{(\beta-1)/2}=n_0^{-0.135}$ on $n_0$. The cloud-scale initial gas inflow rate $\dot{M}_{\rm in}$ and $\eta$ play more important roles, which capture the dynamics of the collapsing cloud evaluated at the density (time) scale $n_0$ ($t_0$). 

Eq.~\ref{r_m_orig} provides a good estimate of the radius of the star-forming disc in the early stage. 
A similar scaling relation $R\propto K^{-1/\beta}(M/\eta)^{\delta/\beta}$ is obtained in \citet[see their eq.~17]{Tan2004} based on angular momentum conservation and analytical spherical collapse solutions \citep{McKee2002,McKee2003}, where $K\equiv P/n^{\gamma_{\rm eff}}\propto Tn^{1-\gamma_{\rm eff}}$ is the entropy parameter. The normalisation of their scaling law is consistent with that in Eq.~\ref{r_m_orig} learned from simulations if the disc mass is similar to the mass of protostars ($\eta\sim 0.5$) and the typical value of $K$ is adopted for infalling gas with $T\sim 300\ \rm K$ and $n\sim 10^{6}\ \rm cm^{-3}$. 
However, this simple power-law scaling can produce unphysically large radii ($\gtrsim 10\ \rm pc$) when $M$ becomes very large ($\gtrsim 10^{4}\ \rm M_\odot$) in the late stage if stellar feedback is inefficient (see Sec.~\ref{sec:mass}). 
Therefore, we further impose an upper limit $R_{\max}$ on the disc radius that can be related to the (initial) cloud size (see Sec.~\ref{sec:mass} below) and introduce the dependence on $\eta$ and $K$:
\begin{align}
    R(M)=\biggl\{R_{\max},(K/K')^{-1/\beta}\left[10^{5\beta}M/(800\eta\ {\rm M_\odot})\right]^{\delta/\beta}\ {\rm AU}\biggr\}\ . \label{r_m}
\end{align}
For simplicity, $K$ is fixed throughout this paper as $K=K'\equiv k_{\rm B}(300\ {\rm K})(10^{6}\ {\rm cm^{-3}})^{1-\gamma_{\rm eff}}$, where $k_{\rm B}$ is the Boltzmann constant, although in reality $K$ depends on the detailed thermo-chemical evolution of the collapsing gas. 
In fact, as discussed below, if $M$ becomes very large ($\gtrsim 10^4\ \rm M_\odot$) before star formation is terminated by feedback, the cloud must undergo violent collapse with high inflow rates $\dot{M}_{\rm in}\gtrsim 0.1\ \rm M_\odot\ yr^{-1}$. 
This usually happens in hotter, atomic-cooling clouds, where $K$ can be higher than that adopted here, leading to smaller discs. {In a companion paper \citep{Gurian2024cloud}}, we develop a semi-analytical model for the thermo-chemistry and collapse dynamics, which can be used to calculate $\dot{M}_{\rm in}$ and $K$ under different conditions of Pop~III star formation \citep[see also][]{Li2021,Sharda2022,Smith2024}. We plan to explore the correlation between $K$ and $\dot{M}_{\rm in}$ in further work using this model\footnote{According to the spherical collapse solution for polytropic gas in \citet[see their eq. 5]{Tan2004}, $K$ weakly correlates with the gas infall rate as $K\propto \dot{M}_{\rm in}^{2\beta/3}\simeq \dot{M}_{\rm in}^{0.49}$. We have verified by numerical experiments that including this weak dependence has minor effects on our results.}.




\subsection{Maximum/final mass of a single star per cloud}
\label{sec:mass}
\begin{figure}
    \centering
    \includegraphics[width=1\columnwidth]{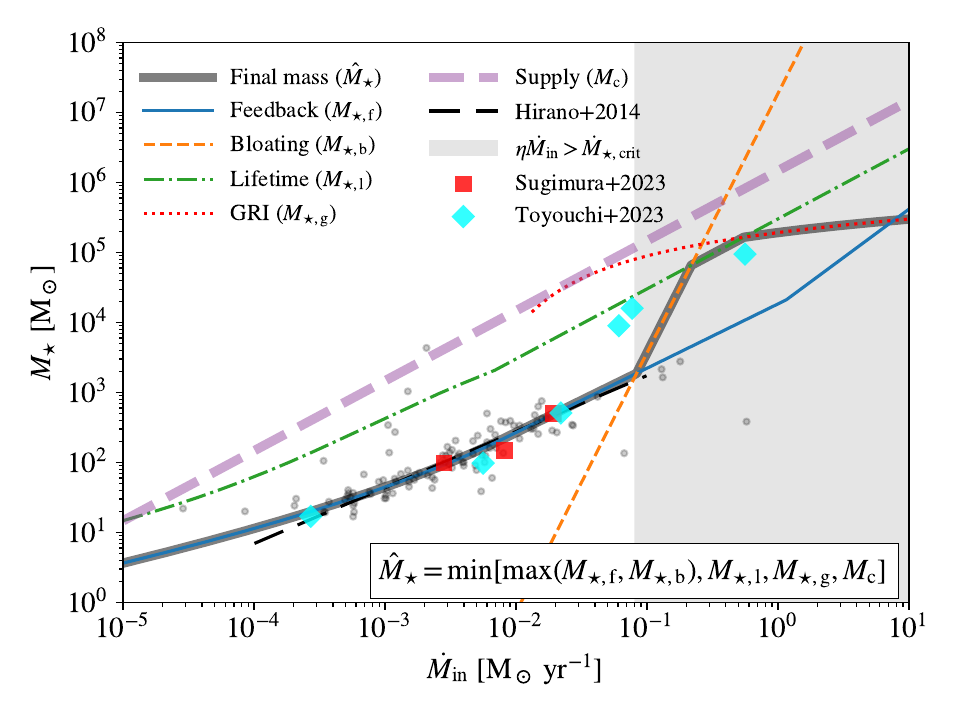}
    \vspace{-20pt}
    \caption{Final stellar mass ($\hat{M}_{\star}$, thick solid), 
    and the characteristic masses regulated by photo-ionization feedback ($M_{\star,\rm f}$, solid), bloating phase ($M_{\star,\rm b}$, dashed), stellar lifetime ($M_{\star,\rm l}$, dash-dotted), GRI ($M_{\star,\rm g}$, dotted), and gas supply ($M_{\rm c}$, thick dashed) as functions of the initial gas inflow rate $\dot{M}_{\rm in}$. 
    The fitting formula of final stellar mass (Eq.~\ref{h14}) based on 2D simulations from \citet[see their fig.~14]{Hirano2014} is shown with the long-dashed line, and the dots show the underlying data. The results of the 3D AMR simulations in \citet[]{Sugimura2023} with rescaled $\dot{M}_{\rm in}$ (see the main text) are denoted by the squares, while those of the 3D spherical simulations in \citet{Toyouchi2023} are shown with the diamonds. The curve for $M_{\star,\rm g}$ is only meaningful when $\dot{M}(M_{\star,\rm g})\gtrsim 0.1\ \rm M_\odot\ yr^{-1}$, corresponding to $\eta\dot{M}_{\rm in}\gtrsim 0.2\ \rm M_\odot\ yr^{-1}$.}
    \label{fig:demo1}
\end{figure}

We first consider the simple case that only one protostar exists in the cloud, so Eqs.~\ref{m_t_final} and \ref{r_m} delineate the (smoothed) median mass and radius evolution of the protostar and circumstellar disc, 
respectively, \textit{governed only by gravity and hydrodynamics}. 
For simplicity, we assume that Eqs.~\ref{m_t_final} and \ref{r_m} hold for $t\gtrsim t_{0}$ until star formation is terminated \textit{quasi-instantaneously} by factors other than gravity and hydrodynamics, such as stellar feedback, stellar collapse/explosion, and gas supply, although in reality they may cause gradual reductions of accretion with long-term deviations from the scaling in Eqs.~\ref{mdot_m} and \ref{m_t_final} \citep{McKee2008}. 
Each terminating effect corresponds to a characteristic maximum/final mass. 
Below, we consider five characteristic masses, which are combined to derive a universal formula for the final mass $\hat{M}_{\star}$ of the central Pop~III star, as illustrated in Fig.~\ref{fig:demo1}.

\begin{enumerate}
    \item \textbf{Final mass regulated by photo-ionization feedback ($M_{\star,\rm f}$):} In our previous work \citep{Gurian2024}, we derive the final mass regulated by photo-ionization feedback assuming that a D-type shock is launched into the cloud once the protostar is thermally relaxed to produce ionizing photons efficiently \citep[][]{Alvarez2006}, and star formation is shut down when the homogenized downstream medium is fully ionized (which marks the evaporation of the cloud). However, the final masses estimated in this way are lower than those found in 2D radiative hydrodynamic simulations \citep{Hirano2014,Hirano2015} by a factor of few. This is mainly caused by the disregard of the circumstellar disc, which can still feed the protostar even if bipolar ionized bubbles have expanded significantly. In fact, it has long been argued that photo-evaporation of the primordial protostellar disc may ultimately set the upper mass scale for Pop~III stars \citep[e.g.,][]{McKee2008,Hosokawa2011,Stacy2012,Fukushima2020}. 
    As shown in \cite{Toyouchi2023}, the feedback-regulated final mass can be estimated from the mass loss rate of the circumstellar disc by photo-evaporation \citep{Tanaka2013}
    \begin{align}
        \dot{M}_{\rm pe}\simeq 0.015\ {\rm M_\odot\ yr^{-1}}\left(\frac{\dot{Q}_{\rm ion}}{10^{52}\ \rm s^{-1}}\right)^{1/2}\left(\frac{R_{\rm pe}}{10^{4}\ \rm AU}\right)^{1/2}\ ,\label{mdot_pe}
    \end{align}
    where $\dot{Q}_{\rm ion}$ is the production rate of ionizing photons (as a function of $M$), and $R_{\rm pe}$ is the physical scale within which the disc suffers from photo-evaporation. Assuming that the accretion rate of the protostar is comparable to the gas inflow rate onto the circumstellar disc, the final mass $M_{\star,\rm f}$ can be obtained by solving $\dot{M}_{\rm pe}=\dot{M}$, above which the disc is rapidly evaporated by ionizing radiation and can no longer feed the protostar. In \citet{Toyouchi2023}, both $\dot{M}$ and $R_{\rm pe}$ are fixed as $\dot{M}=\eta\dot{M}_{\rm in}$ and $R_{\rm pe}=10^{4}\ \rm AU$. Here, we further consider their time evolution, i.e., mass dependence, using the scaling relations in Sec.~\ref{sec:scale}. We set $R_{\rm pe}$ to the circumstellar disc radius given by Eq.~\ref{r_m} (i.e., $R_{\rm pe}\propto M^{1.25}$) and use Eq.~\ref{mdot_m} to derive $\dot{M}(M)$. 
    Finally, to calculate $\dot{Q}_{\rm ion}$, we use the fitting formula from \citet{Schaerer2002} for \textit{non-accreting, non-rotating}\footnote{Pop~III stars are likely born as fast rotators due to rapid accretion of gas with high angular momentum and inefficient magnetic braking during protostar growth \citep[e.g.,][]{Stacy2011,Stacy2013,Hirano2018,Kimura2023}. In the extreme case, Pop~III stars can undergo chemically-homogeneous evolution under efficient mixing triggered by fast rotation. As a result, the star becomes more compact and hotter, such that the production efficiency of ionizing photons $B\equiv \dot{Q}_{\rm ion}/M$ is boosted by a factor of $\sim 2-8$ \citep{Sibony2022}. The final mass regulated by photo-ionization feedback will then be reduced slightly (by up to a factor of 2), as will be shown below that $M_{\star,\rm f}\propto B^{-0.33}$ when $B$ is independent of $M$ for massive stars. } Pop~III stars with $M\sim 5-500\ \rm M_\odot$ combined with the linear relation for massive stars radiating at the Eddington limit \citep{BKL2001,Johnson2012}: 
    \begin{align}
        \dot{Q}_{\rm ion}(M)\simeq \begin{cases}
            10^{a_{0}+a_{1}x+a_{2}x^{2}}\ {\rm s^{-1}}\ ,\quad  &M\le 224\ {\rm M_\odot}\ ,\\
            1.55\times 10^{48+x}\ {\rm s^{-1}}\ ,\quad  &M>224\ {\rm M_\odot}\ ,
        \end{cases}\label{qion}
    \end{align}
    where $x\equiv\log(M/\rm M_\odot)$, $a_{0}=43.61$, $a_{1}=4.9$, and $a_{2}=-0.83$. We assume that the effects of accretion on stellar evolution, such as initial swelling and expansion by radiation pressure enhanced by the accretion shock \citep{Hosokawa2009} become unimportant when the disc is about to diminish under photo-ionization feedback. This assumption is expected to be valid in most cases with $\dot{M}_{\rm in}\gtrsim 10^{-4}\ \rm M_\odot\ yr^{-1}$, where the protostar can reach $M\gtrsim 10\ \rm M_\odot$ within an accretion timescale $t_{\rm acc}(M)=t_{0}(M/M_{0})^{1/\beta}\gtrsim 0.3\ \rm Myr$ that is longer than the Kelvin-Helmholtz (KH) timescale\footnote{We use the protostar evolution model described in appendix~A of \citet{Reinoso2023} to calculate the KH timescale $t_{\rm KH}\equiv t_{\rm KH}(M,\dot{M})$.} $t_{\rm KH}$. Note that disc photo-evaporation only happens after the ionized bubble breaks out into the cloud, while it is shown in the simulations by \citet{Jaura2022} that the bubble can be trapped at the center in the early stage for an extended period $t_{\rm trap}\gtrsim 2\times 10^{4}\ \rm yr$. Here, we assume $t_{\rm acc}>t_{\rm trap}$ such that the initial confinement of ionizing photons does not affect $\dot{M}_{\star,\rm f}$. We find that this assumption is valid as long as $t_{\rm trap}\lesssim t_0=3\times 10^{4}\ \rm yr$.  
    \item \textbf{Maximum mass gained during the feedback-free bloating phase ($M_{\star,\rm b}$):} In the calculation of $M_{\star,\rm f}$, we assume that the protostar is always able to thermally relax and contract, which is only valid for low accretion rates. In fact, the protostar will expand significantly to enter a bloating phase \citep{OmukaiPalla2001,OmukaiPalla2003}, when the accretion rate is above a critical value $\dot{M}_{\star,\rm crit}\sim 0.01-0.04\ \rm M_\odot\ yr^{-1}$ \citep{Hosokawa2013,Haemmerle2018,Herrington2023,Nandal2023}. In this case, the surface temperature becomes as low as $T_{\rm eff}\sim 5000-8000$~K \citep[see fig. 1 in][]{Toyouchi2023}, and the emission of ionizing photons is negligible (with a rate lower than that from a contracted star of the same mass by a factor of $\sim 10^{10}$), which leads to effectively feedback-free evolution. 
    According to Eq.~\ref{mdot_m}, the accretion rate of the protostar slowly decreases with time and will drop below $\dot{M}_{\star,\rm crit}$ at {$t>t_{\rm b}=t_{0}\left(\eta\dot{M}_{\rm in}/\dot{M}_{\star,\rm crit}\right)^{1/(1-\beta)}$ so the maximum mass accreted during the bloating phase is $M_{\star,\rm b}=M_{0}(t_{\rm b}/t_{0})^{\beta}\propto\dot{M}_{\rm in}^{\beta/(1-\beta)}$}. 
    \item \textbf{Maximum mass regulated by the stellar lifetime ($M_{\star, \rm l}$):} Beyond stellar feedback, star formation also terminates when the central protostar runs out of fuel for nuclear fusion in its core and then collapses/explodes. To the first order, this means that the time taken for the protostar to reach a mass $M$, i.e., the accretion timescale $t_{\rm acc}(M)$, cannot exceed the lifetime $t_{\star}(M)$ of a star with the same mass. The maximum mass regulated by the stellar lifetime $M_{\star,\rm l}$ satisfies $t_{\rm acc}(M_{\star,\rm l})=t_{\star}(M_{\star,\rm l})$. It is non-trivial to calculate the lifetimes of accreting massive stars \citep[e.g.,][]{Begelman2010,Johnson2012,Haemmerle2021,Nandal2024acc,Saio2024,Shibata2024}. Here, we simply estimate $t_{\star}$ with (extrapolation of) the fitting formula from \citet{Schaerer2002} for \textit{non-accreting} Pop~III stars (in the mass range $\sim 5-500\ \rm M_\odot$) bound by a lower limit $t_{\star,\min}$ to be calibrated (see below):
    \begin{align}
        t_{\star}(M)=\max\left(t_{\star,\min},10^{b_{0}+b_{1}x+b_{2}x^{2}+b_{3}x^{3}}\ \rm yr\right)\ ,\label{ts}
    \end{align}
    where {$x\equiv\log(M/\rm M_\odot)$,} $b_{0}=9.785$, $b_{1}=-3.759$, $b_{2}=1.413$, and $b_{3}=-0.186$. 
    \item \textbf{Maximum mass regulated by general-relativity instability (GRI, $M_{\star,\rm g}$):} When the accretion rate is very high ($\dot{M}>\dot{M}_{\star,\rm crit,GR}\simeq 0.1\ \rm M_\odot\ yr^{-1}$), the star will collapse by GRI before running out of fuel, and the final stellar mass can be estimated with the fitting formula \citep{Woods2017,Li2021}
    \begin{align}
        M_{\star,\rm GR}=\left[0.83\log\left(\frac{\dot{M}}{\rm M_\odot\ yr^{-1}}\right)+2.48\right]\times 10^{5}\ \rm M_\odot\ ,\label{m_gr}
    \end{align}
    assuming a constant accretion rate \citep[see also][]{Haemmerle2021,Haemmerle2024,Nandal2024acc,Saio2024,Shibata2024}. In our case, the accretion rate $\dot{M}$ decreases \textit{slowly} as $\dot{M}(M)\propto M^{-0.37}\propto t^{-0.27}$ (Eq.~\ref{mdot_m}), and the protostar with a current mass $M$ spends most of its time at accretion rates similar to the current rate $\dot{M}(M)$. We use the current rate to evaluate Eq.~\ref{m_gr} and find the final mass regulated by GRI $M_{\star,\rm g}$ by solving the equation $M=M_{\star,\rm GR}(\dot{M}(M))$ for $M$. 
    \item \textbf{Maximum mass regulated by the gas supply ($M_{\rm c}$):} Finally, the stellar mass cannot exceed the total mass of gas $M_{\rm c}$ available in the collapsing cloud. 
\end{enumerate}

Among the five maximum masses, the first two are closely related to each other. If $M_{\star,\rm b}>M_{\star,\rm f}$, the circumstellar disc will evaporate immediately after the bloating phase ends, and the final mass will be close to $M_{\star,\rm b}$, which we call the bloating mode. On the other hand, if $M_{\star,\rm f}>M_{\star,\rm b}$, 
the protostar can further grow after the bloating phase until it reaches $M_{\star,\rm f}$ to destroy the disc, which we call the feedback-regulated mode. Therefore, if we only consider gravity, hydrodynamics and photo-ionization feedback, the final mass can be well estimated by $\hat{M}_{\star}=\max\left(M_{\star,\rm b},M_{\star,\rm f}\right)$. Considering the other three masses that serve as upper limits, we have 
\begin{align}
    \hat{M}_{\star}=\min\left[\max\left(M_{\star,\rm b},M_{\star,\rm f}\right),M_{\star,\rm l},M_{\star,\rm g},M_{\rm c}\right]\ .\label{mfinal}
\end{align}

There are nine parameters in our single-star model: $\dot{M}_{\rm in}$, $n_0$, $M_{\rm c}$, $K$, $\gamma_{\rm eff}$, $\eta$, $R_{\max}$, $\dot{M}_{\star,\rm crit}$, and $t_{\star,\min}$. The first three reflect the properties of star-forming clouds (and host haloes) on large scales ($\gtrsim1\ \rm pc$). 
The last two parameters are relevant for primordial stellar evolution at small (sub-AU) scales. These two aspects are connected to each other through our model for star-forming discs (Eqs.~\ref{m_t_new}-\ref{r_m}) at intermediate scales governed by $K$, $\gamma_{\rm eff}$, $\eta$, and $R_{\max}$. 

For illustration, below we treat $\dot{M}_{\rm in}$ at $n_0=10^{6}\ \rm cm^{-3}$ as the independent variable and use physically-motivated values for the other parameters. We adopt $\gamma_{\rm eff}=1.09$ for typical primordial collapsing gas under $\rm H_2$ cooling with $K=K'$. The cloud mass $M_{\rm c}$ is associated with $\dot{M}_{\rm in}$ through the collapse timescale $t_{\rm col}$, i.e., $M_{\rm c}=\dot{M}_{\rm in}t_{\rm col}$. As an optimistic estimate, we set the collapse timescale to the free-fall timescale\footnote{The relation between the (cloud-scale) initial gas inflow rate and cloud mass depends on the collapse dynamics and can be generally written as $\dot{M}_{\rm in}=\phi M_{\rm c}/t_{\rm ff}$, where $\phi$ is a numerical parameter of order unity \citep{McKee2002,McKee2003,Tan2004,Omukai2005}. We adopt $\phi=1$ and $n_{\rm crit}=10^{3}\ \rm cm^{-3}$ to obtain conservative (optimistic) estimates of $\dot{M}_{\rm in}$ ($M_{\rm c}$). 
The results of our single-star model are insensitive to the choices of $\phi$ and $n_{\rm crit}$ when $M_{\rm c}$ is large enough given $\phi [n_{\rm crit}/(10^{4}{\ \rm cm^{-3}})]^{1/2}\lesssim 1$. }
$t_{\rm ff}=\sqrt{3\pi/(32Gn_{\rm crit}\mu {m_{\rm H}})}\simeq 1.5\ \rm Myr$ for a critical density $n_{\rm crit}=10^{3}\ \rm cm^{-3}$, beyond which $\rm H_{2}$ cooling saturates in minihaloes \citep{ABN2002,BCL2002,Gurian2024} and $\rm H_{2}$ formation becomes efficient in atomic-cooling haloes \citep{Sugimura2024}. {This density also marks the onset of collapse by runaway cooling when a core with a enclosed mass larger than the Bonnor-Ebert mass first emerges \citep{Gurian2024cloud}. }
Given the cloud mass and the critical density, we estimate the initial cloud size as $R_{\rm c}=[3M_{\rm c}/(4\pi\mu n_{\rm crit}{m_{\rm H}})]^{1/3}$, which then sets the upper limit of the circumstellar disc radius $R_{\max}=R_{\rm c}$ in Eq.~\ref{r_m}. We further fix $t_{\star,\min}$ using the critical accretion rate $\dot{M}_{\star,\rm crit,GR}\simeq 0.1\ \rm M_\odot\ yr^{-1}$ above which GRI collapse is expected to occur \citep[see their fig.~4]{Woods2017}. To be specific, we require that $\dot{M}=\dot{M}_{\star,\rm crit,GR}$ when $M_{\star,\rm l}=M_{\star,\rm g}$, which gives $t_{\star,\min}\simeq 1.2\ \rm Myr$. We adopt an intermediate value $\eta=0.5$ based on the finding $\eta\sim 0.4-0.6$ from simulations \citep{Sakurai2016,Toyouchi2023}. This means that half of the infalling gas goes into the protostar system, while the rest constitutes the star-forming disc (or leaves the star-forming cloud via ejected protostars and outflows). Following \citet{Toyouchi2023}, we use $\dot{M}_{\star,\rm crit}= 0.04\ \rm M_\odot\ yr^{-1}$ by default to obtain conservative estimates of $\hat{M}_{\star}$. 
In reality, these input parameters can have non-trivial correlations with $\dot{M}_{\rm in}$ and each other and show a strong dependence on environmental factors (e.g., LW radiation and streaming motion between baryons and dark matter). For simplicity, we ignore such correlations in this section and explore their possible impact on the results in Appendix~\ref{apdx:details}.


Fig.~\ref{fig:demo1} 
shows the characteristic masses $M_{\star,\rm b}$, $M_{\star,\rm f}$, $M_{\star,\rm l}$, $M_{\star,\rm g}$, $M_{\rm c}$, and the final mass $\hat{M}_{\star}$ as functions of $\dot{M}_{\rm in}$. 
The transition from the feedback-regulated mode ($\hat{M}_{\star}=M_{\star,\rm f}$) to the bloating mode ($\hat{M}_{\star}=M_{\star,\rm b}$) occurs almost exactly at $\dot{M}_{\rm in}=\dot{M}_{\star,\rm crit}/\eta=0.08\ \rm M_\odot\ yr^{-1}$, corresponding to $\hat{M}_{\star}\sim 2000\ \rm M_\odot$. 
Approaching this transition point, when the production rate of ionizing photons is proportional to stellar mass as $\dot{Q}_{\rm ion}(M)=BM$ for massive stars with $M\gtrsim 200\ \rm M_\odot$, {$M_{\star,\rm f}$ can be derived analytically for $R<R_{\max}$: 
\begin{align}
    &\dot{M}(M_{\star,\rm f})=\dot{M}_{\rm pe}(M_{\star,\rm f})\ ,\\
    &(\eta \dot{M}_{\rm in})^{\frac{1}{\beta}}(\beta M_{\star,\rm f}/t_0)^{\frac{\beta-1}{\beta}}=C(BM_{\star,\rm f})^{\frac{1}{2}}K^{-\frac{1}{2\beta}}(M_{\star,\rm f}/\eta)^{\frac{\delta}{2\beta}}\ ,\\
    &(\eta \dot{M}_{\rm in})^{\frac{1}{\beta}}(\beta/t_0)^{\frac{\beta-1}{\beta}}M_{\star,\rm f}^{\frac{\beta-1}{\beta}}=CB^{\frac{1}{2}}K^{-\frac{1}{2\beta}}\eta^{-\frac{\delta}{2\beta}}M_{\star,\rm f}^{\frac{\beta+\delta}{2\beta}}\ ,\\
    &M_{\star,\rm f}^{\frac{\beta-2-\delta}{2\beta}}=(\beta/t_0)^{\frac{1-\beta}{\beta}}CB^{\frac{1}{2}}K^{-\frac{1}{2\beta}}\eta^{-\frac{\delta+2}{2\beta}}\dot{M}_{\rm in}^{-\frac{1}{\beta}}\ ,\\
    &M_{\star,\rm f}=(\beta/t_0)^{2(\beta-1)\epsilon}C^{-2\beta\epsilon}B^{-\beta\epsilon}K^{\epsilon}\eta^{(\delta+2)\epsilon}\dot{M}_{\rm in}^{2\epsilon}\ .
\end{align}
Here we have simplified the final expression by defining $\epsilon\equiv 1/(2+\delta-\beta)\simeq 0.46$ and used Eqs.~\ref{mdot_m}, \ref{m_0}, \ref{r_m}, and \ref{mdot_pe} to obtain the second line given the constant $C\equiv 0.015\ {\rm M_\odot\ yr^{-1}}\left[10^{-56}K'^{1/\beta}10^{5\delta}(800\ {\rm M_\odot})^{-\delta/\beta}\ {\rm s}\right]^{1/2}$. A quasi-linear relation between $\hat{M}_{\star}=M_{\star,\rm f}$ and $\dot{M}_{\rm in}$ is found as $M_{\star,\rm f}\propto B^{-\beta\epsilon}K^{\epsilon}\eta^{(\delta+2)\epsilon}\dot{M}_{\rm in}^{2\epsilon}\simeq B^{-0.33}K^{0.46}\eta^{1.33}\dot{M}_{\rm in}^{0.92}$. On the other hand, in the bloating mode, the final stellar mass increases rapidly with $\dot{M}_{\rm in}$ as $\hat{M}_{\star}=M_{\star,\rm b}\propto\dot{M}_{\rm in}^{\beta/(1-\beta)}\simeq \dot{M}_{\rm in}^{2.7}$} before it reaches the limits placed by the stellar lifetime and GRI. The gas supply ($M_{\rm c}$) never becomes a limiting factor, 
and we always have $t_{\rm acc}>t_0$.

\begin{figure}
    \centering
    \includegraphics[width=1\columnwidth]{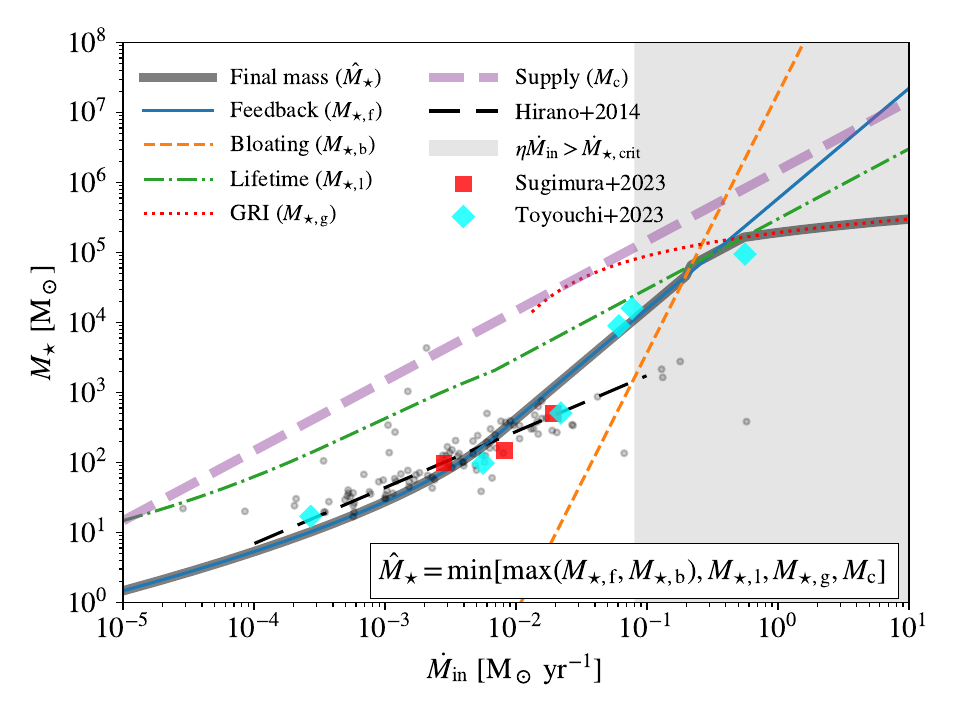}
    \vspace{-20pt}
    \caption{Same as Fig.~\ref{fig:demo1} but assuming a fixed scale $R_{\rm pe}=10^{4}$~AU for disc photo-evaporation following \citet{Toyouchi2023}. As expected, the simulation results from \citet[denoted by the diamonds]{Toyouchi2023} are better reproduced here compared with the case in Fig.~\ref{fig:demo1} with disc size evolution $R_{\rm pe}\propto M^{1.25}$.}
    \label{fig:demo2}
\end{figure}

In the feedback-regulated regime, our results ($M_{\star,\rm f}$) agree perfectly with the fitting formula based on 2D simulations (for $\dot{M}_{\rm in}\sim 10^{-4}-0.1\ \rm M_\odot\ yr^{-1}$) from \citet[see their fig.~14]{Hirano2014}
\begin{align}
    M_{\star}=100\ {\rm M_\odot}\left(\frac{\dot{M}_{\rm in}}{2.8\times 10^{-3}\ \rm M_\odot\ yr^{-1}}\right)^{0.8}\ ,\label{h14}
\end{align}
and the results of the 3D simulations with adaptive mesh refinement (AMR) over Cartesian coordinates in \citet[]{Sugimura2023}, which produce 2 to 4 stars per cloud in the end rather than a single star. Here, we have applied a correction factor $\simeq 3.7$ to the initial gas inflow rates in \citet[see their table~1]{Sugimura2023} to be consistent with the definition of $\dot{M}_{\rm in}$ in \citet{Hirano2014}\footnote{Underlying the fitting formula Eq.~\ref{h14} from \citet{Hirano2014}, the initial gas inflow rate $\dot{M}_{\rm in}$ is measured {at the moment the \textit{central} hydrogen density reaches $n_{\rm H,cen}=10^{12}\ \rm cm^{-3}$ and at the radius where the ratio between enclosed mass and Bonnor–Ebert} mass reaches the maximum, corresponding to a density $n\sim 10^{6}\ \rm cm^{-3}$ (see their figs.~11 and 13) consistent with our definition of $\dot{M}_{\rm in}$ for $n_0=10^{6}\ \rm cm^{-3}$. When $\dot{M}_{\rm in}$ is evaluated at an earlier stage with a smaller central density $n_{\rm H,cen}=10^{7}\ \rm cm^{-3}$ in the 2D simulations of \citet{Hirano2015} as well as in the 3D simulations of \citet{Sugimura2023}, a best-fit $M_{\star}=250\ {\rm M_\odot}[\dot{M}_{\rm in}/(2.8\times 10^{-3}\ \rm M_\odot\ yr^{-1})]^{0.7}$ is obtained with a similar slope but a higher normalisation, because $\dot{M}_{\rm in}$ is lower compared with the $n_{\rm H,cen}=10^{12}\ \rm cm^{-3}$ case. The two fits are consistent with each other given the relation $\dot{M}_{\rm in}(n_{\rm H,cen}=10^{12}\ {\rm cm^{-3}})=3.7\dot{M}_{\rm in}(n_{\rm H,cen}=10^{7}\ {\rm cm^{-3}})$, which is used to correct the inflow rates reported in \citet{Sugimura2023}.}. The final mass is underestimated by our model at $\dot{M}_{\rm in}\lesssim 10^{-4}\ \rm M_\odot$ compared with the results of two simulations from \citet{Hirano2014} in this regime where the initial KH contraction phase likely becomes important. This so-called low-mass mode of Pop~III star formation \citep{Stacy2014} is expected to be rare. 


We also compare our predictions with the results of the 3D single-star simulations with spherical coordinates in \citet[see their table~3]{Toyouchi2023}, finding a good agreement for $\dot{M}_{\rm in}\lesssim 0.02\ \rm M_\odot\ yr^{-1}$. This shows that our model accurately captures the median evolution of Pop~III protostars and circumstellar discs in the feedback-regulated regime at least for $\dot{M}_{\rm in}\lesssim 0.02\ \rm M_\odot\ yr^{-1}$. 
However, two simulations in \citet{Toyouchi2023} with $\dot{M}_{\rm in}\sim 0.06-0.08\ \rm M_\odot\ yr^{-1}$ obtain higher masses ($\sim 10^{4}\ \rm M_\odot$) than our results ($\hat{M}_\star\sim 2000\ \rm M_\odot$) and those from the 2D simulations in \citet{Hirano2014} under similar conditions. This implies that the final mass can vary greatly from system to system around the critical inflow rate $\sim \dot{M}_{\star,\rm crit}/\eta$ required to trigger the bloating phase, likely due to the deviation from the scaling relations (Eqs.~\ref{m_t_new}-\ref{r_m}) in the (stochastic) evolution of individual systems, which we will discuss in Sec.~\ref{sec:caveats} and Appendix~\ref{apdx:scal}. In the lifetime/GRI-regulated regime, our model reproduces the result of \citet{Toyouchi2023} for $\dot{M}_{\rm in}\sim 0.6\ \rm M_\odot\ yr^{-1}$ within a factor of 2, showing a slight overestimation\footnote{The final mass for $\dot{M}_{\rm in}\sim 0.6\ \rm M_\odot\ yr^{-1}$ in \citet[see their LWH-10 model]{Toyouchi2023} is obtained by assuming constant mass growth for a duration of $0.5\ \rm Myr$ using the average accretion rate in the first 0.2~Myr covered by the simulation. Therefore, it should be regarded as a lower limit if the gas inflow can be sustained beyond 0.5~Myr.}.  

For comparison, Fig.~\ref{fig:demo2} 
shows the results assuming a fixed scale $R_{\rm pe}=10^{4}$~AU for disc photo-evaporation following \citet{Toyouchi2023}. 
As expected, the predicted $\hat{M}_{\star}$-$\dot{M}_{\rm in}$ relation is generally consistent with the simulation data from \citet{Toyouchi2023}. 
The final mass regulated by photo-ionization feedback $M_{\star,\rm f}$ is overestimated (by up to a factor of $\sim 10$) at $\dot{M}_{\rm in}\sim 0.02-0.1\ \rm M_\odot\ yr^{-1}$ compared with the fit from \citet{Hirano2014}\footnote{Similarly, $M_{\star,\rm f}$ is overestimated by a factor of $\sim 10$ in \citet[see their eq.~27]{Li2021}, where smaller values of $R_{\rm pe}$ are adopted based on the gravitational influence radii of protostars. }. In this way, the feedback-regulated mode is also important for $\dot{M}_{\rm in}\sim 0.08-0.2\ \rm M_\odot\ yr^{-1}$, 
and the bloating mode becomes rare. 
However, for $\dot{M}_{\rm in}\lesssim 0.002\ \rm M_\odot\ yr^{-1}$, $M_{\star,\rm f}$ is slightly lower for $R_{\rm pe}=10^{4}$~AU compared with the fit from \citet{Hirano2014} and our results with $R_{\rm pe}\propto M^{1.25}$. This indicates that the evolution of $R_{\rm pe}$ is crucial for the accurate modelling of photo-ionization feedback and calculation of $\hat{M}_\star$. 

\subsection{Effects of fragmentation/multiplicity}
\label{sec:multiplicity}

\begin{figure*}
    \includegraphics[width=0.8\textwidth]{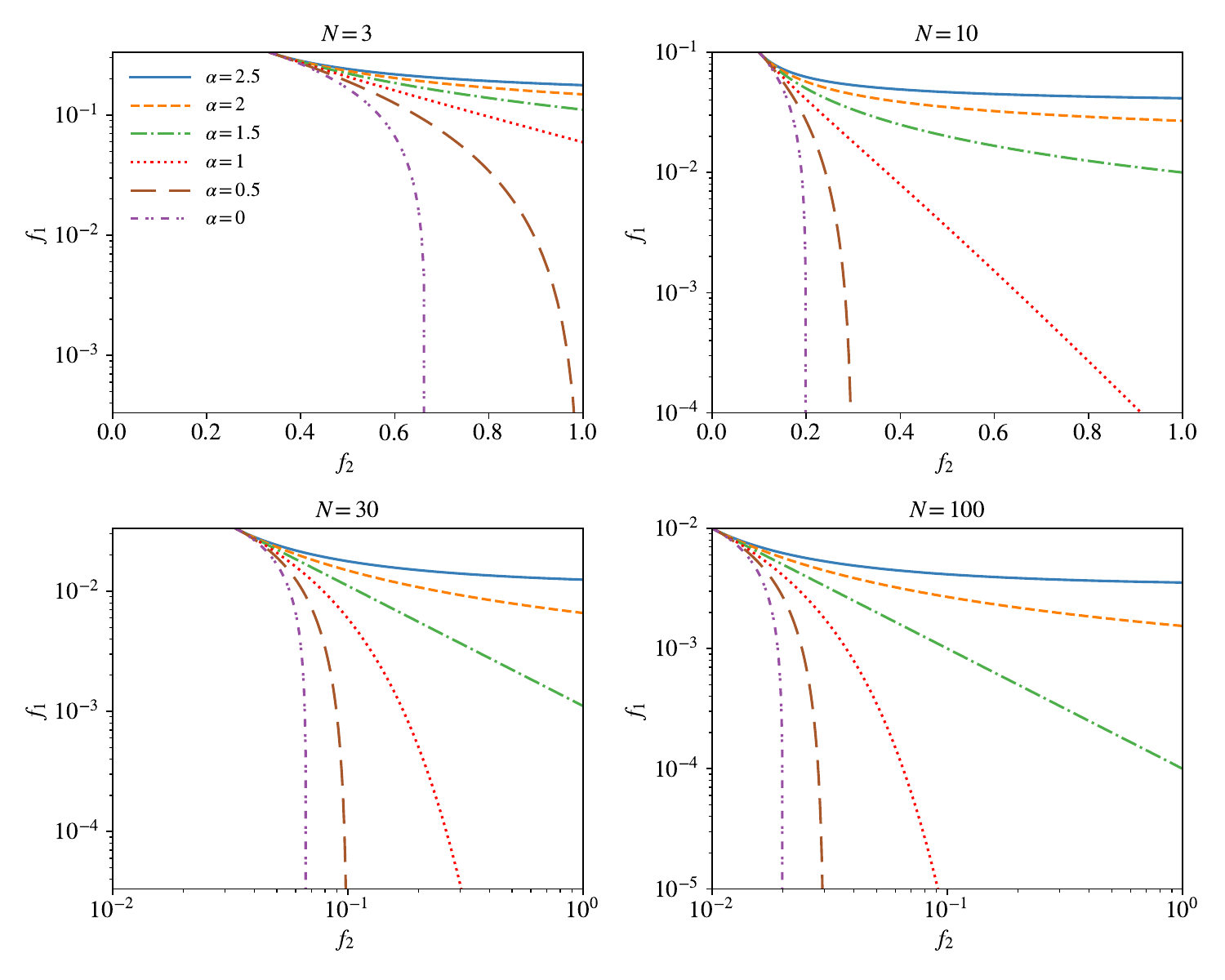}
    \vspace{-15pt}
    \caption{Relation between $f_{1}$ and $f_{2}$ for $\alpha=2.5$ (solid), 2 (dashed), 1.5 (dash-dotted), 1 (dotted), 0.5 (long-dashed), and 0 (dash-dot-dotted) given $N=3$ (top-left), 10 (top-right), 30 (bottom-left), and 100 (bottom-right).}
    \label{fig:f1}
\end{figure*}

In the previous subsection, we derive the final stellar mass under the assumption that only one protostar forms per cloud. However, in the last two decades, small-scale simulations have shown that fragmentation is common in primordial star-forming clouds/discs \citep{Bromm2013,Haemmerle2020,Klessen2023}, producing multiple protostars that undergo complex evolution and interactions (accretion, migration, gravitational scatters, and mergers). There is also evidence for multiplicity in Pop~III star formation from stellar archaeology data \citep{Hartwig2023}. However, the final fates of the protostars are still in debate due to computational limitations (e.g., resolution, time coverage, volume, and treatment of feedback). For instance, in the dense discs formed under strong gas inflows ($\dot{M}_{\rm in}\gtrsim 0.1\ \rm M_\odot\ yr^{-1}$), gaseous dynamical friction can drive rapid inward migrations of fragments/protostars to merge with one central object that dominates the total stellar mass, converging to the single-star scenario, or feed a small number of SMSs and many low-mass stars \citep{Inayoshi2014,Boekholt2018,Suazo2019,Chon2020,Latif2021,Woods2021,Woods2024,Prole2023bh,Reinoso2023,Schleicher2023,Kiyuna2024}. 
In general, the interplay between fragmentation and feedback has profound implications for the formation efficiency and IMF of Pop~III stars, similar to the case of Pop~I/II star formation \citep[e.g.,][]{Menon2024}.

In light of this, we generalize the single-star model in Sec.~\ref{sec:mass} to follow the growth and feedback of multiple protostars per cloud and explore the effects of multiplicity on the final total stellar mass. 
We assume that the cloud has modest turbulence at the onset of collapse due to inefficient cooling \citep{Chon2021} and only forms one star-forming disc that fragments by disc instability. 
We only follow the median evolution of the mass distribution of \textit{surviving} protostars and ignore transient protostars that are promptly lost in mergers or ejections\footnote{Although ejections are commonly seen in simulations \citep[e.g.,][]{Prole2022res,Raghuvanshi2023,Reinoso2023}, their effects on the final total stellar mass are still unclear. Since protostars can hardly grow once ejected from the star-forming cloud, the mass carried by ejected stars is expected to be negligible compared with the final mass of the remaining star cluster. However, the radiative feedback from ejected stars may also regulate star formation. We defer the inclusion of ejected stars to future work.}. 
As a heuristic attempt, we consider a simple phenomenological multiplicity model under the following two assumptions.
\begin{enumerate}
    \item The mass distribution of protostars follows a power law $dN/dm\propto m^{-\alpha}$. 
    \item The fraction of the total accretion rate $\dot{m}/\dot{M}$ that feeds a protostar is identical to the fraction of total stellar mass $m/M$ occupied by the protostar. Therefore, if the number of protostars is constant, the mass distribution of protostars undergoes self-similar evolution: The shape (i.e., power-law index) remains unchanged and the boundaries increase proportionally with the total stellar mass $M$. 
\end{enumerate}
Here, the second assumption is motivated by the quasi-linear relation between $\dot{m}/\dot{M}$ and $m/M$ found in the (magneto-) hydrodynamic simulations from \citet{Sharda2020} when the accretion rate is measured over a timescale $\gtrsim 0.01 t_{\rm KH}$ in the early stage with $M\lesssim50\ \rm M_\odot$. 
{Interestingly, this quasi-linear relation between accretion rate and mass appears to be an intermediate case compared with the extreme regimes of tidal lobe accretion \citep[$\dot{m}\propto m^{2/3}$,][]{Bonnell2001} and Bondi-Hoyle accretion \citep[$\dot{m}\propto m^{2}$,][]{Bondi1944}. According to the theory of competitive accretion in star clusters \citep{Bonnell2001}, tidal lobe accretion is expected to happen in the early stage when the cluster mass/potential is dominated by gas \citep{Bonnell2001comp}, while Bondi-Hoyle accretion takes over once stars dominate the potential and are virialized. For Pop~III star formation, the masses of stars and gas are comparable in the star-forming disc ($\eta\sim 0.5$), so it is reasonable that the accretion rate-mass relation has a power-law slope $\sim 1$ between $2/3$ and 2. 

With self similarity,} the protostar mass distribution is determined by three parameters $\alpha$, $f_{1}$, and $f_{2}$, where the last two parameters define the lower and upper bounds as $m_{\min}=f_{1}M$ and $m_{\max}=f_{2}M$. These parameters are related to the number of protostars $N$ through 
\begin{align}
    \frac{1}{N}=\frac{1}{M}\frac{\int_{f_{1}M}^{f_{2}M}m^{1-\alpha}dm}{\int_{f_{1}M}^{f_{2}M}m^{-\alpha}dm}=\frac{\int_{f_{1}}^{f_{2}}f^{1-\alpha}df}{\int_{f_{1}}^{f_{2}}f^{-\alpha}df}\ ,
\label{f1}
\end{align}
where $f\equiv m/M$ given $m$ the mass of individual protostars. 
We treat $\alpha$ and $f_{2}$ as free parameters and associate $f_{1}$ to $N$, such that $N$ becomes the third parameter of our model. 
Given $\alpha$, $f_{2}$, and $N$, we first solve Eq.~\ref{f1} to obtain $f_{1}(N,f_{2},\alpha)$, as shown in Fig.~\ref{fig:f1}. Then we distribute $N$ bins $f_{i}$ in the range $f\in(f_{1},f_{2})$ according to a power law with index $\alpha$ and apply small corrections to ensure $\sum_{i}^{N}f_{i}=1$. The masses and accretion rates of individual protostars are simply $m_{i}=f_{i}M$ and $\dot{m}_{i}=f_{i}\dot{M}$ given the total stellar mass $M$ and accretion rate $\dot{M}(M)$ of the protostar cluster (Eqs.~\ref{mdot_m} and \ref{m_0}). Here, we assume that the evolution of $\dot{M}$ with $M$ (governed by cloud-scale collapse) is completely determined by $\dot{M}_{\rm in}$ and not affected by the fragmentation process before feedback kicks in, since recent simulations have shown that {the growth of total stellar mass is insensitive to the number of protostars at least in the early stage without feedback \citep[see, e.g.,][]{Wollenberg2020,Liu2021binary,Prole2022res}}. In this way, \textit{the effects of fragmentation on the final total stellar mass only manifest via its interplay with feedback}, as discussed below.

Once the relation between $M$ and the masses and accretion rates of individual protostars is known, we generalize the analysis in Sec.~\ref{sec:mass} to multiple protostars. We first absorb the effects of the bloating phase into the calculation of ionizing photon production rates, in which we reduce the value from Eq.~\ref{qion} by a factor of $10^{10}$ when the accretion rate is above $\dot{M}_{\star,\rm crit}$ \citep[see fig.~4 in][]{Toyouchi2023}. We derive the total disc photo-evaporation rate under the assumption that 
all protostars reside in a central region much smaller than the star-forming disc and contribute to the evaporation of the disc effectively as one point source. This central concentration may occur by inward migrations from dynamical friction \citep{Riaz2023}. 
In this case, we simply replace $\dot{Q}_{\rm ion}$ in Eq.~\ref{mdot_pe} with the summation of the rates $\dot{q}_{{\rm ion},i}$ from individual protostars and still use the radius of the entire disc from Eq.~\ref{r_m} with $R_{\max}=R_{\rm c}$ to estimate the disc evaporation scale 
\begin{align}
    \dot{M}_{\rm pe}\simeq 0.015\ {\rm M_\odot\ yr^{-1}}\left(\frac{\sum_{i}^{N}\dot{q}_{{\rm ion},i}}{10^{52}\ \rm s^{-1}}\right)^{1/2}\left[\frac{R(M)}{10^{4}\ \rm AU}\right]^{1/2}\ .\label{mdot_pe_whole}
\end{align}
An alternative scenario with distributed feedback is explored in Appendix~\ref{apdx:fdbk}.

Next, we consider the limits on the total stellar mass placed by the lifetime and GRI of the most massive protostar in the cluster (with the index $i=N$), assuming that once this star ends its life, gas inflows will be shut down either by a supernova explosion or by black hole accretion feedback. To be specific, we now use $m_{N}=f_{N}M$ rather than $M$ to evaluate the stellar lifetime $t_{\star}$ formula (Eq.~\ref{ts}) and compare $m_{N}$ with the GRI mass from Eq.~\ref{m_gr} (using $\dot{m}_{N}=f_{N}\dot{M}$ instead of $\dot{M}$). The final total stellar mass $\hat{M}_{\star}$ is given by the maximum value of $M$ that satisfies
\begin{align}
\begin{split}
    &\dot{M}_{\rm pe}(M,f_{i})\le \dot{M}\ ,\\
    &t_{\rm acc}(M)=t_{0}(M/M_{0})^{1/\beta}\le t_{\star}(f_{N}M)\ ,\\
    &f_{N}M\le M_{\star,\rm GR}(f_{N}\dot{M})\ ,
\end{split}\label{m_f_frag}
\end{align}
and again capped by the cloud mass $M_{\rm c}$. 

To demonstrate the effects of multiplicity on $\hat{M}_{\star}$, we explore the parameter space\footnote{An alternative parametrisation of multiplicity focusing on the minimum mass of Pop~III stars is investigated in Appendix~\ref{apdx:param}.} of $\alpha$, $f_{2}$, and $N$ with $3\times4=12$ examples while fixing $\dot{M}_{\star,\rm crit}=0.04\ \rm M_\odot\ yr^{-1}$ and $\eta=0.5$. Here, we assume that $N$ is constant (i.e., independent of $M$) for simplicity and consider the concentrated feedback scenario (Eq.~\ref{mdot_pe_whole}). In practice, we impose an upper limit $N_{\max}$ on $N$ corresponding to $f_{1}(N_{\max},f_{2},\alpha)=0.1\ {\rm M_\odot}/M$ when evaluating Eq.~\ref{m_f_frag} to avoid the production of objects below the minimum mass $\sim 0.1\ \rm M_\odot$ of Pop~III protostars \citep[see their fig.~7]{Greif2011}. Therefore, $N$ is not strictly constant in the model, and the input value of $N$ as a parameter should be regarded as the optimistic target value. These choices ($\dot{M}_{\star,\rm crit}=0.04\ \rm M_\odot\ yr^{-1}$, $\eta=0.5$, constant $N$, and concentrated feedback) together with the assumptions of $R_{\rm pe}\propto M^{1.25}$ and $\gamma_{\rm eff}=1.09$ are adopted as the default setup unless otherwise noted. More detailed results, 
including those with different prescriptions/assumptions and choices of parameters for fragmentation, mass-size scaling relation, and photo-ionization feedback are shown in Appendix~\ref{apdx:details}.

\begin{figure}
    \centering
    \includegraphics[width=1\columnwidth]{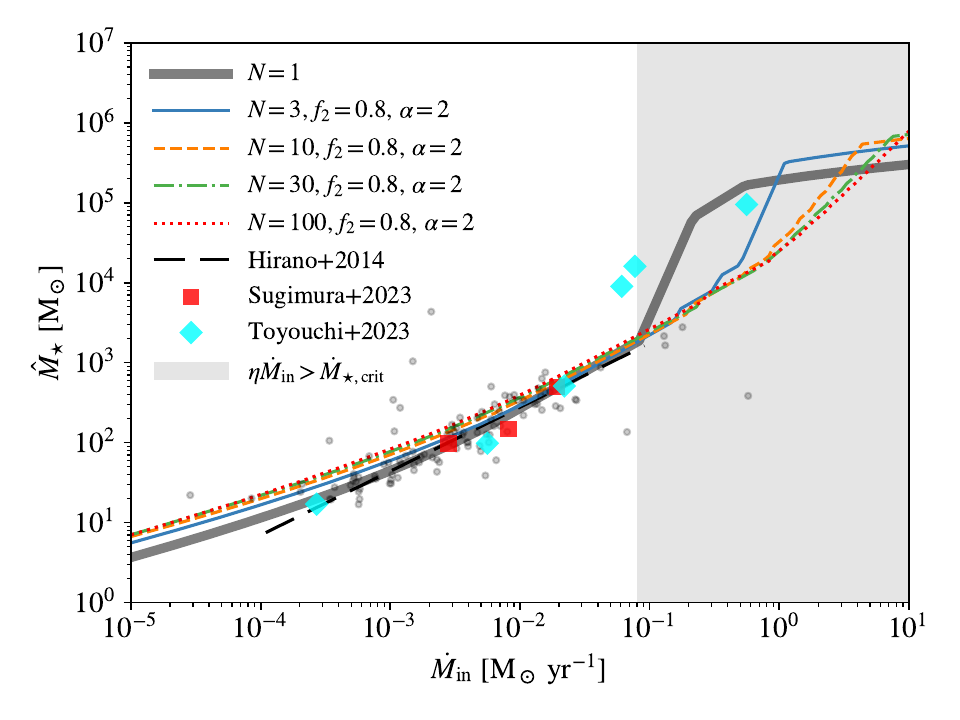}
    \vspace{-20pt}
    \caption{Final total stellar mass as a function of initial gas inflow rate 
    for $N=3$ (thin solid), 10 (dashed), 30 (dash-dotted), and 100 (dotted) with $f_{2}=0.8$ and $\alpha=2$. 
    The results for the single-star model ($N=1$) is shown with the thick solid curve for comparison. The fitting formula of final stellar mass (Eq.~\ref{h14}) based on 2D (single-star) simulations from \citet[see their fig.~14]{Hirano2014} is shown with the long-dashed line, and the dots show the underlying data. The results of the 3D AMR simulations in \citet{Sugimura2023} are denoted by the squares, while those of the 3D spherical simulations in \citet{Toyouchi2023} are shown with the diamonds.}
    \label{fig:m_N}
\end{figure}

\begin{figure}
    \centering
    \includegraphics[width=1\columnwidth]{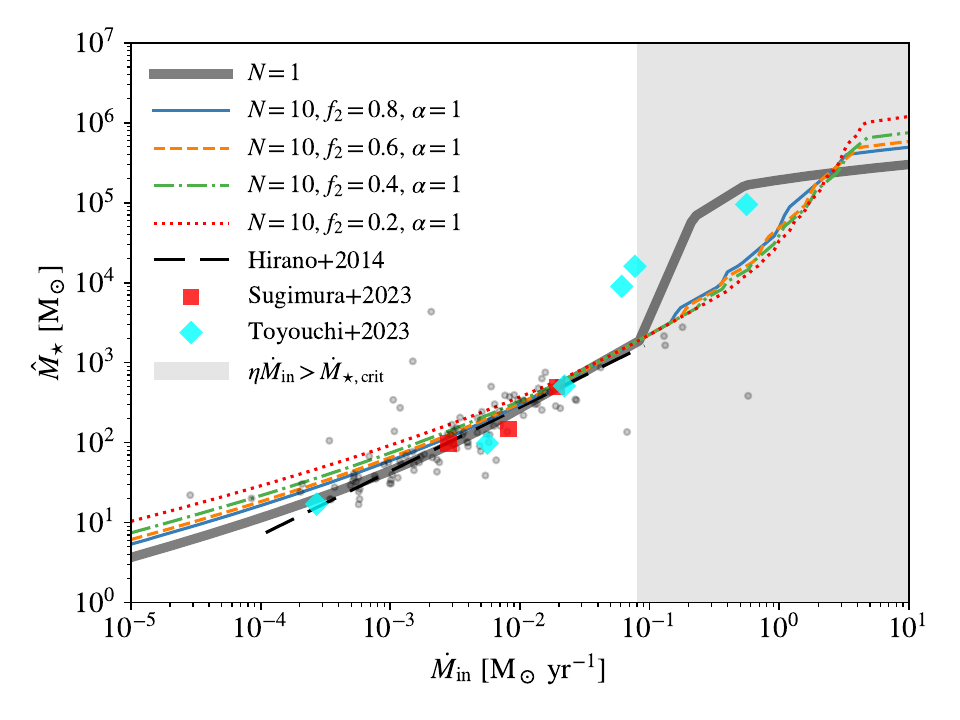}
    \vspace{-20pt}
    \caption{
    Same as Fig.~\ref{fig:m_N} but for $f_{2}=0.8$ (thin solid), 0.6 (dashed), 0.4 (dash-dotted), and 0.2 (dotted) with $N=10$ and $\alpha=1$. }
    \label{fig:m_f2}
\end{figure}

\begin{figure}
    \centering
    \includegraphics[width=1\columnwidth]{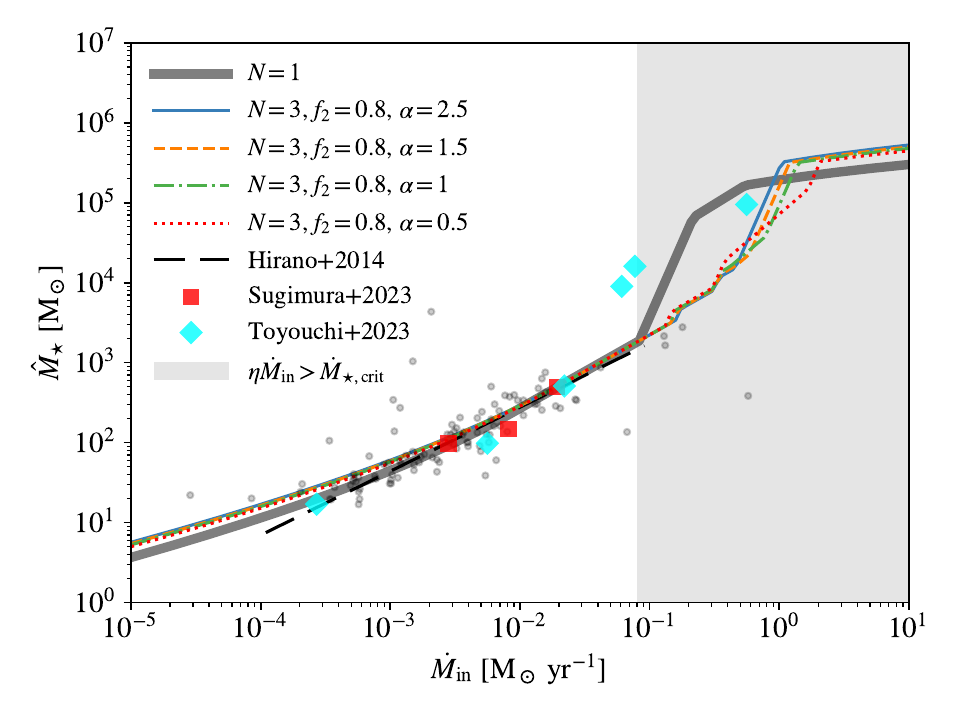}
    \vspace{-20pt}
    \caption{
    Same as Fig.~\ref{fig:m_N} but for $\alpha=2.5$ (thin solid), 1.5 (dashed), 1 (dash-dotted), and 0.5 (dotted) with $N=3$ and $f_{2}=0.8$.}
    \label{fig:m_alpha}
\end{figure}

Fig.~\ref{fig:m_N} shows the relation between $\hat{M}_{\star}$ and $\dot{M}_{\rm in}$ for $N=3$, 10, 30, and 100 given $f_{2}=0.8$ and $\alpha=2$. 
Note that $f_{1}$ decreases with $N$ (Fig.~\ref{fig:f1}), low-mass stars make up a larger fraction of the total stellar mass given a larger $N$, which are less efficient at producing ionizing photons compared with massive stars if we do not consider the bloating phase. Therefore, as $N$ increases, photo-ionization feedback is weaker in the low-accretion rate regime $\eta\dot{M}_{\rm in}\lesssim \dot{M}_{\star,\rm crit}$ without bloating, leading to higher $\hat{M}_{\star}$. The effect is stronger for smaller $\dot{M}_{\rm in}$. On the contrary, for $\eta\dot{M}_{\rm in}\gtrsim \dot{M}_{\star,\rm crit}$, the transition from the no-bloating feedback-regulated regime ($\hat{M}_{\star}\lesssim 2000\ \rm M_\odot$, where all stars have contracted before star formation is terminated by feedback) to the lifetime/GRI-regulated regime ($\hat{M}_{\star}\gtrsim 10^{5}\ \rm M_\odot$) is smoother and delayed as $N$ increases, leading to smaller $\hat{M}_{\star}$ in the intermediate stage (corresponding to the bloating mode in the single-star model). This is because the full bloating phase, where all protostars have accretion rates above $\dot{M}_{\star,\rm crit}$, is more difficult to trigger when we distribute the total accretion rate among multiple protostars. Even if the most massive stars have entered the bloating phase, low-mass stars can still contract and produce ionizing photons efficiently under low accretion rates. The fraction of stars that experience bloating when star formation terminates increases with $\dot{M}_{\rm in}$ until it reaches unity. However, $\hat{M}_{\star}$ increases again with $N$ in the GRI-regulated regime, because the GRI mass limit (Eq.~\ref{m_gr}) is applied to the most massive star, whose mass fraction in the cluster is lower with higher $N$. 

Fig.~\ref{fig:m_f2} shows the results for $f_{2}=0.8$, 0.6, 0.4, and 0.2 given $N=10$ and $\alpha=1$. Similarly to trends with $N$, $\hat{M}_{\star}$ decreases with $f_{2}$ for $\eta\dot{M}_{\rm in}\lesssim \dot{M}_{\star,\rm crit}$ and in the GRI-regulated regime, but increases with $f_{2}$ in the intermediate regime. This is because the importance of massive stars simply increases with $f_{2}$, resulting in stronger feedback if there is no bloating but meanwhile making the bloating phase and GRI easier to occur. Besides, since $f_{1}$ increases when $f_{2}$ is smaller (Fig.~\ref{fig:f1}), the transition to the GRI-regulated regime is generally sharper. 

Fig.~\ref{fig:m_alpha} shows the results for $\alpha=2.5$, 1.5, 1, and 0.5 given $N=3$ and $f_{2}=0.8$. Here, the dependence on $\alpha$ is generally weaker compared to the trends in $N$ and $f_{2}$. The biggest effect is that the full bloating phase occurs at lower $\dot{M}_{\rm in}$ when the protostar mass distribution is more bottom-heavy with higher $\alpha$ because $f_{1}$ decreases with $\alpha$ when the input values of $N$ and $f_{2}$ are fixed (Fig.~\ref{fig:f1}).

\section{Discussion}
\label{sec:dis}

\subsection{Implications for the cluster mass distribution and IMF of Pop~III stars in quasar progenitor haloes}
\label{sec:mdis}

As an example, we apply our analytical model for Pop~III star formation to the primordial star-forming clouds in the (main) progenitor haloes of high-$z$ luminous quasar host galaxies and derive the mass distributions of Pop~III clusters ($M_\star$) and stars ($m_\star$). The (cloud-scale initial) gas inflow rates $\dot{M}_{\rm in}$ of such clouds are calculated by \citet{Li2021} with merger trees targeting typical luminous quasar host haloes with masses $M_{\rm h}=10^{12}\ \rm M_\odot$ 
and a co-moving number density\footnote{We derive the number density of haloes with $M_{\rm h}\gtrsim 10^{12}\ \rm M_\odot$ at $z=6$ from the halo mass function in $\Lambda$CDM with the \textsc{colossus} package \citep{Diemer2018} using the \citet{Tinker2008} model and the \texttt{planck15} set of cosmological parameters \citep{PlanckCollaboration2016}. See \url{https://bdiemer.bitbucket.io/colossus/lss_mass_function.html}.} of $n_{\rm h}\sim 2.3\times 10^{-6}\ \rm cMpc^{-3}$ at $z=6$, 
corresponding to $\sim 4\sigma$ over-dense peaks that only occupy a small fraction $\lesssim 6\times 10^{-5}$ of cosmic volume in the standard $\Lambda$CDM cosmology \citep{PlanckCollaboration2016}. 
They use a one-zone model to follow the cloud collapse until $n= 10^{6}\ \rm cm^{-3}$ to predict the gas inflow rate, consistent with our definition of $\dot{M}_{\rm in}$ for $n_0=10^6\ \rm cm^{-3}$. 
In addition to standard thermo-chemistry of collapsing primordial gas, they also model the effects of Lyman-Werner (LW) radiation from nearby star-forming galaxies, dynamical heating by halo mergers, and streaming motion between baryons and dark matter, which can delay Pop~III star formation to more massive (atomic-cooling) clouds/haloes with higher gas inflow rates, favoring the formation of SMSs and heavy BH seeds \citep[see, e.g.,][for reviews]{Inayoshi2020,Haemmerle2020,Volonteri2021,Wise2023,Regan2024}. 

\begin{figure}
    \centering
    \includegraphics[width=1\columnwidth]{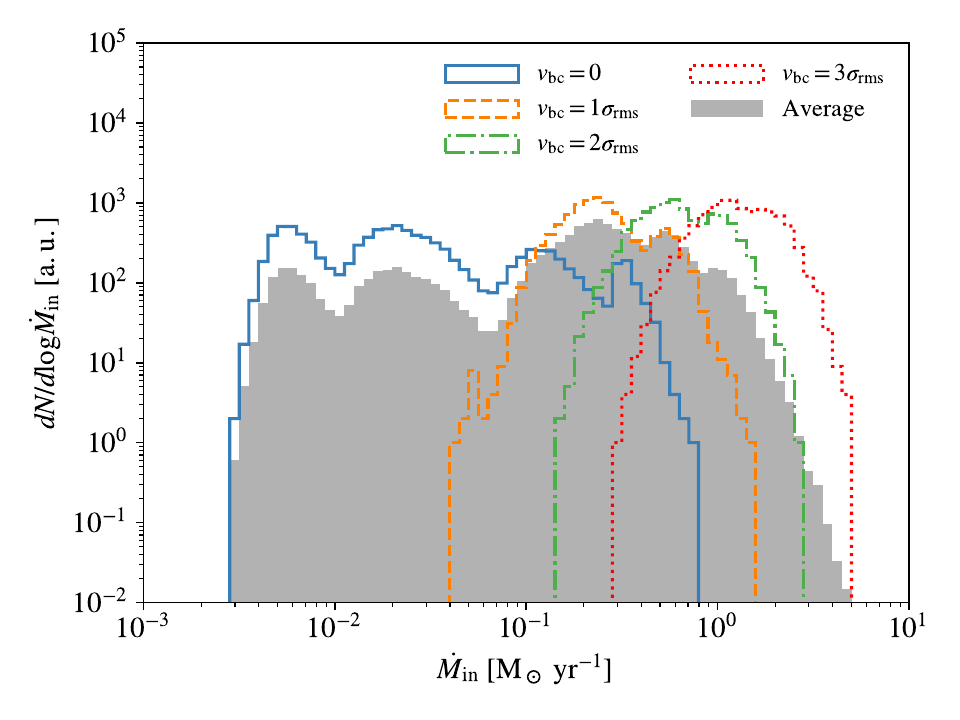}
    \vspace{-20pt}
    \caption{Gas inflow rate distribution of primordial star-forming clouds in high-$z$ luminous quasar progenitor haloes predicted by \citet[see their fig.~6]{Li2021} for $v_{\rm bc}=0$ (solid), 1 (dashed), 2 (dash-dotted) and $3\sigma_{\rm rms}$ (dotted), each with 10000 realisations. The shaded region shows the cosmic average distribution estimated by numerical integration of the distributions for the four $v_{\rm bc}$ values using the trapezium rule, given the probability density function $p(v_{\rm bc})$ of $v_{\rm bc}$ (Eq.~\ref{pvbc}), following the same normalisation, i.e., $\int (dN/d\dot{M}_{\rm in})d\dot{M}_{\rm in}=10000$. Here, the number counts for different $v_{\rm bc}$ models are not directly comparable, as they are not normalized to reflect $p(v_{\rm bc})$ but to have the same total number of haloes/clouds.}
    \label{fig:mdot_dis}
\end{figure}

Here we consider the $\dot{M}_{\rm in}$ distributions 
predicted by \citet{Li2021} for four values of baryon-dark matter streaming velocity $v_{\rm bc}=0$, 1, 2, and $3\sigma_{\rm rms}$ (where $\sigma_{\rm rms}$ is the cosmic root-mean-squared velocity), as shown in Fig.~\ref{fig:mdot_dis}.  
The $\dot{M}_{\rm in}$ distribution for $v_{\rm bc}=0$ covers a broad range $\dot{M}_{\rm in}\sim 0.003-1\ \rm M_\odot\ yr^{-1}$ with multiple peaks corresponding to haloes of different assembly histories and LW background intensities \citep[see their fig.~6]{Li2021}, and the distribution is narrower and dominated by higher $\dot{M}_{\rm in}$ as $v_{\rm bc}$ increases, reaching up to $\dot{M}_{\rm in}\sim 5\ \rm M_\odot\ yr^{-1}$ for $v_{\rm bc}=3$. 
We further construct the $\dot{M}_{\rm in}$ distribution in the cosmic (volume) average case by numerically integrating the results for the four $v_{\rm bc}$ values over the probability density function of $v_{\rm bc}$ \citep{Tseliakhovich2011,Fialkov2014,Schauer2019}
\begin{align}
    p(v_{\rm bc})=\left(\frac{3}{2\pi\sigma_{\rm rms}}\right)^{3/2}4\pi v_{\rm bc}^{2}\exp\left(-\frac{3v_{\rm bc}^{2}}{2\sigma_{\rm rms}^{2}}\right)\ ,\label{pvbc}
\end{align}
using the trapezium rule. To be specific, the results for $v_{\rm bc}/\sigma_{\rm rms}=0$, 1, 2, and 3 are combined with the weights $w_{0}=0.5F(0,1)\simeq 0.304$, $w_{1}=0.5[F(0,1)+F(1,2)]\simeq 0.496$, $w_{2}=0.5[F(1,2)+F(2,3)]\simeq 0.196$, and $w_{3}=0.5F(2,3)+F(3,\infty)\simeq 0.004$, respectively, where the function $F$ is defined as $F(x,y)\equiv \int_{x\sigma_{\rm rms}}^{y\sigma_{\rm rms}}p(v_{\rm bc})dv_{\rm bc}$. Since we only have 4 bins of $v_{\rm bc}$, the numerical integration is only meant to capture the general shape of the cosmic average distribution, while the detailed features in the integrated distribution may be numerical artifacts, which can be smoothed out given a finer grid of $v_{\rm bc}$. 

\begin{figure}
    \centering
    \includegraphics[width=1\columnwidth]{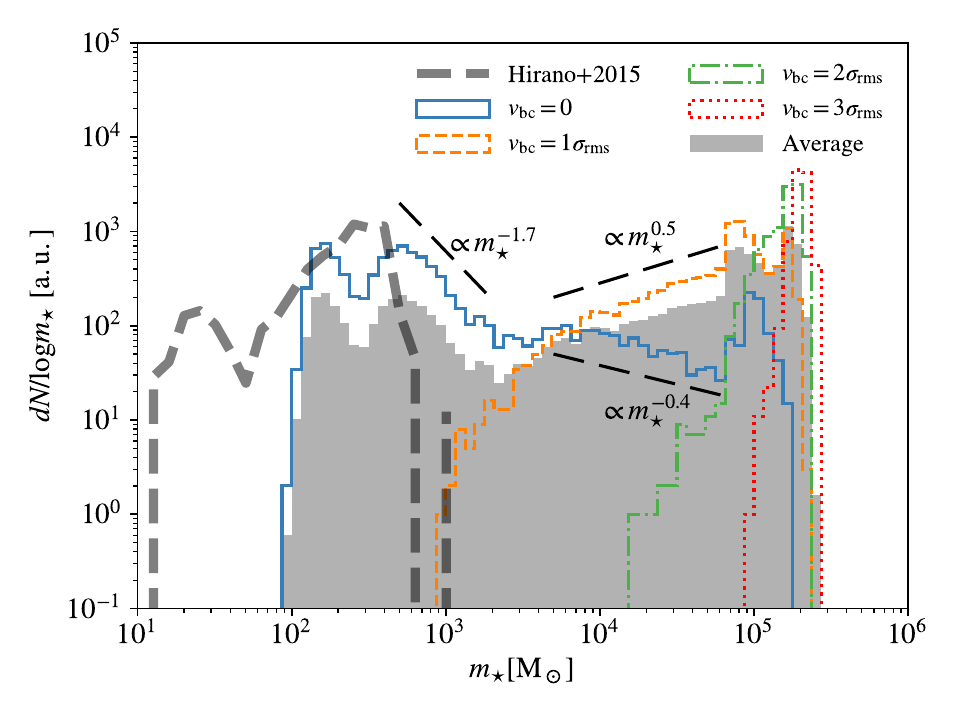}
    \vspace{-20pt}
    \caption{Mass distribution of Pop~III stars formed in high-$z$ luminous quasar progenitor haloes based on the $\dot{M}_{\rm in}$ data (Fig.~\ref{fig:mdot_dis}) from \citet[]{Li2021} for $v_{\rm bc}=0$ (solid), 1 (dashed), 2 (dash-dotted), $3\sigma_{\rm rms}$ (dotted), and the cosmic average (shaded), assuming a single star per cloud/halo. The distribution approximately follows a (broken) power-law form in the range $m_\star\sim 500-7\times 10^{4}\ \rm M_\odot$ for $v_{\rm bc}=0$ and the cosmic average, as indicated by the long-dashed lines. The mass distribution of typical Pop~III stars formed in unbiased regions with $v_{\rm bc}=0$ from \citet{Hirano2015} is shown with the thick dashed curve for comparison. These distributions are normalized such that the total number of haloes is 10000 in each case.}
    \label{fig:mc_dis_N1}
\end{figure}

Fig.~\ref{fig:mc_dis_N1} shows the resulting mass distributions of Pop~III stars in the single-star model ($N=1$, $m_{\star}=M_{\star}$) under the default setup for $v_{\rm bc}=0$, 1, 2, and $3\sigma_{\rm rms}$ as well as the cosmic average. For comparison, we also plot the mass distribution of Pop~III stars formed in typical minihaloes in unbiased regions with $v_{\rm bc}=0$ from the 2D (single-star) simulations in \citet{Hirano2015}. Some statistical properties of these mass distributions are summarized in Table~\ref{tab:my_label}. 
In the simple case with no streaming motion ($v_{\rm bc}=0$), the mass distribution spans a wide range $m_{\star}\sim 100-2\times 10^{5}\ \rm M_\odot$ with a complex shape caused by the multimodal distribution of $\dot{M}_{\rm in}$. There are two low-mass peaks around $m_{\star}\sim 160$ and $500\ \rm M_\odot$ and a high-mass peak at $m_\star\sim 10^{5}\ \rm M_\odot$ that contains a small fraction ($\sim 7\%$) of stars. In between the second low-mass peak and the high mass peak, the distribution approximately follows a broken power-law form with $dN/d\log m_\star\propto m_\star^{-1.7}$ at $m_\star\sim 500-2000\ \rm M_\odot$ and $dN/d\log m_\star\propto m_\star^{-0.4}$ at $m_\star\sim 5000-7\times 10^{4}\ \rm M_\odot$. Although the majority ($\sim 80\%$) of primordial star-forming clouds in high-$z$ luminous quasar progenitor haloes form ordinary Pop~III stars with $m_\star\sim 100-2000\ \rm M_\odot$ (in the feedback-regulated regime), a non-negligible fraction ($N_{\rm SMS}\sim 14\%$) of clouds produce SMSs with $m_\star>10^{4}\ \rm M_\odot$ that can become heavy BH seeds. Besides, a small fraction ($\sim 8\%$) of clouds produce VMSs with $m_\star\sim 2000-9000\ \rm M_\odot$ that can explain the high nitrogen abundances observed in several high-$z$ galaxies by JWST \citep{Nandal2024}. 

Similarly to our case with $v_{\rm bc}=0$, the mass distribution predicted by \citet{Hirano2015} is also dominated by ordinary Pop~III stars ($m_\star\sim 100-1000\ \rm M_\odot$) with a peak at $m_\star\sim 300\ \rm M_\odot$. However, it has a sharp cutoff at $\sim 1000\ \rm M_\odot$ in the lack of enhanced environmental effects (halo mergers and LW radiation) that can produce high $\dot{M}_{\rm in}$ and more massive Pop~III stars in high-$z$ luminous quasar progenitor haloes. \citet{Hirano2015} also obtain a small ($\sim 12\%$) fraction of Pop~III stars below $100\ \rm M_\odot$, mostly formed in HD cooling haloes at $z\lesssim 15$, while such cases are extremely rare in high-$z$ luminous quasar progenitor haloes, since Pop~III star formation mostly occur at $z\sim 15-50$ in the relevant biased regions \citep{Li2021}. The masses of Pop~III stars are significantly boosted by higher streaming velocities between baryons and dark matter, such that most ($\gtrsim 90\%$) stars have $m_\star>10^{4\ (5)}\ \rm M_\odot$ for $v_{\rm bc}=1\ (2)\sigma_{\rm rms}$, while the number of VMSs is reduced. With the contributions of such high-$v_{\rm bc}$ regions, the cosmic average mass function of Pop~III stars is dominated ($\sim 70\%$) by SMSs, showing a very top-heavy power law $dN/d\log m_\star\propto m_\star^{0.5}$ at $m_\star\sim 3000-7\times 10^{4}\ \rm M_\odot$ followed by two higher-mass peaks that account for half of the entire population. These differences indicate that the environmental effects at halo and cosmic scales strongly regulate the masses of Pop~III stars. In general, the broad mass distributions predicted by our analytical model are consistent with the picture emerging from recent numerical simulations that the mass spectrum of BH seeds from the first stars is a continuum \citep{Li2023,Regan2024}.

\begin{table}
    \centering
    \caption{Statistics of Pop~III clusters and stars in high-$z$ luminous quasar progenitor haloes. Column 1: streaming motion velocity, where `average' denotes the cosmic average case constructed by numerical integration based on Eq.~\ref{pvbc}. Columns 2-4: multiplicity parameters (Sec.~\ref{sec:multiplicity}). Column 5: fraction of haloes hosting massive clusters with $M_{\star}>10^{4}\ \rm M_\odot$. Column 6: average number per halo of VMSs in the mass range $m_\star\sim 2000-9000\ \rm M_\odot$ as sources of large nitrogen yields \citep{Nandal2024}. Column 7: average number per halo of SMSs with $m_\star>10^{4}\ \rm M_\odot$ as progenitors of heavy BH seeds. 
    Column 8: average number per halo of extremely massive stars above $10^5\ \rm M_\odot$ whose BH remnants are able to grow rapidly \citep{Regan2024}.}
    \begin{tabular}{cccccccc}
        \toprule
         $v_{\rm bc}$ & $N$ & $f_{2}$ & $\alpha$ & $f_{\rm MC}$ & $N_{\rm VMS}$ & $N_{\rm SMS}$ & $N_{>5}$ \\
        \bottomrule
         0 & 1 & - & - & 0.1384 & 0.0825 & 0.1384 & 0.0336 \\
        $1\sigma_{\rm rms}$ & 1 & - & - & 0.9232 & 0.0625 & 0.9232 & 0.2622 \\
        $2\sigma_{\rm rms}$ & 1 & - & - & 1 & 0 & 1 & 0.9347 \\
        $3\sigma_{\rm rms}$ & 1 & - & - & 1 & 0 & 1 & 0.9999 \\
        Average & 1 & - & - & 0.6998 & 0.0561 & 0.6998 & 0.3271 \\
        \hline
        Average & 3 & 0.8 & 2 & 0.3237 & 0.9262 & 0.4149 & 0.0505 \\
        Average & 10 & 0.8 & 2 & 0.2482 & 0.8781 & 0.0698 & 5.4e-5 \\
        Average & 30 & 0.8 & 2 & 0.2457 & 0.5775 & 0.0301 & 1.1e-6\\
        Average & 100 & 0.8 & 2 & 0.2659 & 0.4561 & 0.0193 & 0\\
        \hline
        Average & 10 & 0.8 & 1 & 0.2915 & 0.9699 & 0.2474 & 0.0011 \\
        Average & 10 & 0.6 & 1 & 0.2919 & 1.0592 & 0.1655 & 0.0004 \\
        Average & 10 & 0.4 & 1 & 0.2718 & 0.9872 & 0.1077 & 6.4e-5\\
        Average & 10 & 0.2 & 1 & 0.2180 & 0.7562 & 0.0406 & 4.3e-5\\
        \hline
        Average & 3 & 0.8 & 2.5 & 0.3235 & 0.8560 & 0.4735 & 0.0589\\
        Average & 3 & 0.8 & 1.5 & 0.3245 & 0.9620 & 0.3771 & 0.0385\\
        Average & 3 & 0.8 & 1 & 0.3260 & 0.9779 & 0.3690 & 0.0220\\
        Average & 3 & 0.8 & 0.5 & 0.3353 & 0.9118 & 0.4424 & 0.0033\\
        \bottomrule
    \end{tabular}
    \label{tab:my_label}
\end{table}

Given the number density $n_{\rm h}\sim 2.3\times 10^{-6}\ \rm cMpc^{-3}$ of typical high-$z$ luminous quasar host haloes (with $M_{\rm h}\sim 10^{12}\ \rm M_\odot$ at $z\sim 6$), 
the number density of heavy ($\gtrsim 10^{4}\ \rm M_\odot$) BH seeds from Pop~III SMSs formed in their main progenitors is $n_{\rm seed}=N_{\rm SMS}n_{\rm h}\sim 1.6\times 10^{-6}\ \rm cMpc^{-3}$ (averaging over regions with different $v_{\rm bc}$). 
This should be regarded as a \textit{lower limit} of the abundance of Pop~III heavy BH seeds in high-$z$ luminous quasar progenitor haloes, as Pop~III formation can occur in multiple branches of a merger tree beyond the main branch. 
Nevertheless, this lower limit already accounts for $\sim 11\%$ of the number density $n_{\rm quasar}\sim 1.4\times 10^{-5}\ \rm cMpc^{-3}$ of luminous quasars with bolometric luminosities $L_{\rm bol}\gtrsim 10^{46}\ \rm erg\ ^{-1}$ at $z\sim 6$ inferred from JWST observations\footnote{The fact that $n_{\rm h}<n_{\rm quasar}$ implies that the majority of quasars newly discovered by JWST reside in less massive ($M_{\rm h}<10^{12}\ \rm M_\odot$) haloes at $z\sim 6$ than those considered in \citet{Li2021} which are expected to host the most massive BHs with $m_{\rm BH}\gtrsim 10^{9}\ \rm M_\odot$ \citep{Wyithe2006}. One needs to model Pop~III SMS formation for a broader halo mass range to fully evaluate the roles played by Pop~III stars in seeding SMBHs, which we defer to future work.} \citep[see their fig.~8]{Akins2024}. 
Given Eddington ratios and duty cycles of the order of unity \citep{Fragione2023}, such heavy BH seeds from Pop~III SMSs (formed at $z\gtrsim 15$) can grow to $m_{\rm BH}\gtrsim 10^{8}\ \rm M_\odot$ by $z\sim 6$, powering luminous quasars with $L_{\rm bol}\sim L_{\rm Edd}\simeq 1.2\times 10^{38}\ {\rm erg\ s^{-1}}(m_{\rm BH}/\rm M_\odot)\gtrsim 10^{46}\ \rm erg\ ^{-1}$. 
In fact, about half of the Pop~III SMSs (and their BH remnants) have masses above $10^{5}\ \rm M_\odot$, which is the minimum BH mass required to activate efficient growth found in recent simulations \citep{Regan2024}.  
A detailed investigation for the {formation \citep[e.g.,][]{Haemmerle2021,Haemmerle2024,Cammelli2024,Coughlin2024,Nagele2024,Nandal2024acc,Saio2024,Shibata2024} and growth \citep[e.g.,][]{Johnson2016,Latif2020,Toyouchi2021,Inayoshi2022,Hu2022,Park2022bh,Li2023,Lupi2023,Massonneau2023,Sassano2023,Bhowmick2024,Jeon2023,Jeon2024} of BH seeds from Pop~III SMSs is beyond the scope of this paper}.
In general, Pop~III SMSs can make a significant contribution to the (heavy) BH seed population underlying high-$z$ luminous quasars, at least under the single-star assumption.

\begin{figure}
    \centering
    \includegraphics[width=1\columnwidth]{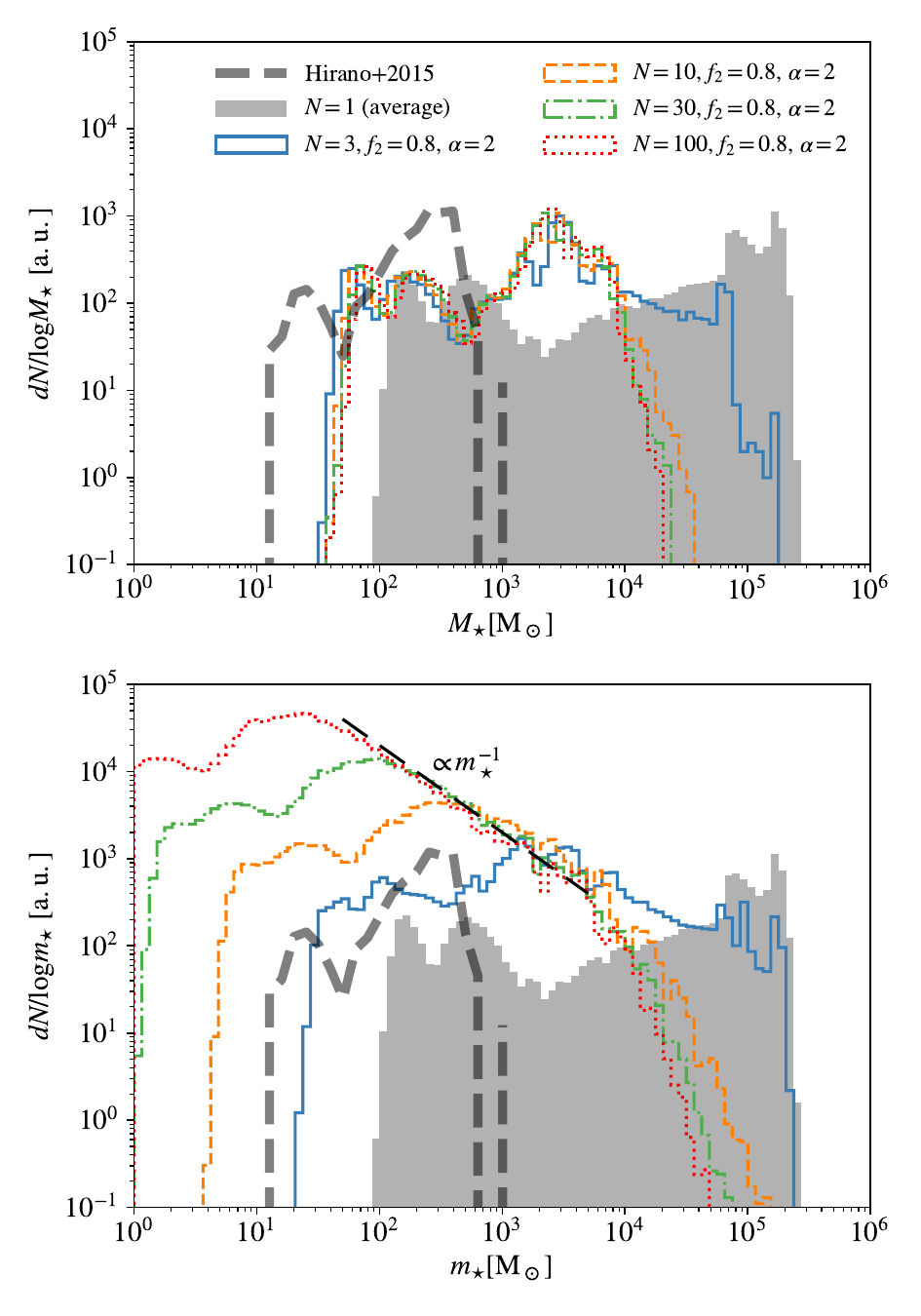}
    \vspace{-20pt}
    \caption{Mass distributions of Pop~III clusters (top) and stars (bottom) formed in high-$z$ luminous quasar progenitor haloes for $N=3$ (thin solid), 10 (dashed), 30 (dash-dotted), and 100 (dotted) with $f_{2}=0.8$ and $\alpha=2$. Here, the cosmic average distributions are constructed from the results with different values of $v_{\rm bc}$ using Eq.~\ref{pvbc}. The mass distribution in the single-star model ($N=1$) is shown with the shaded region. For $N\ge 10$, the mass distribution of stars has a power-law shape similar to that assumed in the cloud/disc-scale multiplicity model ($\alpha=2$) in a mass range that is broader for a larger $N$, as illustrated by the long-dashed line. The mass distribution of typical Pop~III stars formed in unbiased regions with $v_{\rm bc}=0$ from \citet{Hirano2015} is shown with the thick dashed curve, which is normalized to have a total number of haloes of 10000 consistent with the other distributions.}
    \label{fig:m_dis_N}
\end{figure}

Next, we explore the effects of multiplicity at the smaller scales of star-forming clouds. The 12 multiplicity models shown in Sec.~\ref{sec:multiplicity} and listed in Table~\ref{tab:my_label} (with $N>1$) are applied to the $\dot{M}_{\rm in}$ datasets from \citet{Li2021}. For conciseness, we focus on the dependence on $N$ as the most important multiplicity parameter, considering the cosmic average case for $v_{\rm bc}$. The mass distributions of Pop~III clusters and stars with $N=3$, 10, 30, and 100 given $f_{2}=0.8$ and $\alpha=2$ are shown in Fig.~\ref{fig:m_dis_N}. The corresponding statistics are summarized in Table~\ref{tab:my_label}, which also include the results for the other 8 models varying $f_{2}$ and $\alpha$. As expected, fragmentation generally suppresses the formation of massive Pop~III clusters with $M_{\star}\gtrsim 10^{4}\ \rm M_\odot$, reducing their number fraction from $f_{\rm MS}\sim 70\%$ in the single-star model to $f_{\rm MS}\sim 25-32\%$ for $N\sim 3-100$, as the full bloating phase becomes rare in the presence of low-mass slowly-accreting stars. However, in the $N=3$ case, the maximum total stellar mass increases to $M_\star\sim 4\times 10^{5}\ \rm M_\odot$ because the collapse of the most massive star by GRI is delayed. 
Interestingly, the Pop~III cluster masses predicted by our model remain below the upper limit $M_{\star,\max}\sim 4\times 10^{5}-7\times 10^{5}\ \rm M_\odot$ \citep{Liu2020} estimated from the non-detection of Pop III-dominated galaxies in the Hubble Frontier Fields at $z\sim 6-9$ with a limiting rest-frame UV magnitude of -13.5 \citep{Bhatawdekar2021} using the stellar population synthesis code \textsc{yggdrasil} \citep{Zackrisson2011}. 
The decrease of $f_{\rm MS}$ with $N$ slows down at $N\gtrsim 10$ and is even reversed for $N\gtrsim 30$, because the ionizing power is lower when there are more (low-mass) stars for a fixed total stellar mass in the no-bloating feedback-regulated regime. 
In fact, there is always a significant fraction $f_{\rm MC}\sim 22-34\%$ of haloes hosting massive clusters for the 12 models considered here covering $N\sim 3-100$, $f_2\sim 0.2-0.8$, and $\alpha\sim 2.5-0.5$. This indicates that massive star clusters are common outcomes of primordial star formation in over-dense regions, whose $N$-body dynamics has crucial implications on the formation and evolution of merging Pop~III BBHs and their gravitational wave signatures \citep{Wang2022,Liu2023sc,Mestichelli2024}.

The number of SMSs is also reduced as $N$ increases, showing a sharp drop from $N=3$ to $N=10$, down to $N_{\rm SMS}\sim 0.02$ per halo for $N=100$. The corresponding number density of heavy BH seeds is $n_{\rm seed}\simeq 9.8\times 10^{-7}\ (4.5\times 10^{-8})\ \rm cMpc^{-3}$ for $N=3\ (100)$, lower than the observed number density of $z\sim 6$ luminous ($L_{\rm bol}\gtrsim 10^{46}\ \rm erg\ s^{-1}$) quasars by a factor of $\sim 15\ (300)$. Besides, given $N\gtrsim 3$, most ($\gtrsim 80\%$) SMSs have $m_\star\lesssim 10^{5}\ \rm M_\odot$, which may not be massive enough to produce rapidly growing BH seeds \citep{Regan2024}. This indicates that fragmentation can significantly reduce the efficiency of heavy BH seed formation in Pop~III star-forming clouds. 
On the other hand, formation of VMSs with $m_\star\sim 2000-9000\ \rm M_\odot$ as potential sources of strong nitrogen enrichment \citep{Nandal2024} is significantly enhanced by fragmentation. The number of such VMSs is $N_{\rm VMS}\sim 0.5-0.9$ per halo for $N\sim 3-100$ compared with $N_{\rm VMS}\sim 0.06$ for $N=1$. This implies that understanding the effects of fragmentation/multiplicity on the final products of primordial star formation is essential for interpreting the peculiar observational signatures of high-$z$ galaxies/quasars unveiled by JWST \citep{Adamo2024}. Note that here we focus on the natal population of stars/BHs, while VMSs, SMSs, and heavy BH seeds can further form and grow via run-away collisions in massive clusters \citep[e.g.,][]{Sakurai2017,Boekholt2018,Wang2022}. 
Last but not least, when $N$ is larger, the shape of the overall IMF of Pop~III stars is increasingly influenced by the protostar mass distribution assumed in the multiplicity model and extends to lower masses (down to $m_\star\sim 1\ \rm M_\odot$ for $N\gtrsim 30$). The mass distribution of stars follows a power-law form $dN/d\log m_\star\propto m_\star^{-1}$ consistent with the input IMF at the cloud/disc scales ($\alpha=2$) from $m_\star\sim 5000\ \rm M_\odot$ to $m_\star\sim 500/100/50\ \rm M_\odot$ for $N=10/30/100$, while in the $N=3$ case, the mass distribution of stars resembles more the cluster mass distribution regulated by larger-scale physics. 

Compared with $N$, the effects of $f_{2}$ and $\alpha$ are weaker. The most obvious trend is that $N_{\rm SMS}$ decreases when $f_{2}$ is lower (see the third section in Table~\ref{tab:my_label}). Indeed, in our default multiplicity model, beyond $N$, it is mainly $f_{2}$ (rather than $\alpha$) that characterizes the oligarchy of the most massive objects from competitive accretion, which is important for SMS formation. It will be shown in Appendix~\ref{apdx:param} that $\alpha$ can have stronger effects under an alternative parametrisation of multiplicity. 

We end this section with some brief comments on the Pop~III SFE in high-$z$ luminous quasar progenitor haloes, defined as $f_{\star}\equiv M_\star/M_{\rm gas}$ given the total gas mass $M_{\rm gas}$ in the halo at the moment of star formation. A broad range $f_\star\sim 0.001-0.1$ is obtained regardless of the choices of multiplicity parameters. In the single-star model, the efficiency is generally higher in more massive haloes with larger $M_\star$. The typical values are $f_\star\sim 0.004$ for ordinary Pop~III star formation ($M_\star< 2000\ \rm M_\odot$) in minihaloes with $M_{\rm h}\sim 2\times 10^{5}-3\times 10^{6}\ \rm M_\odot$ and $f_{\star}\sim 0.02$ for SMS formation ($M_\star > 10^{4}\ \rm M_\odot$) in atomic-cooling haloes ($M_{\rm h}\gtrsim 10^{7}\ \rm M_\odot$). When fragmentation is included, $f_\star$ hardly changes in minihaloes but is significantly reduced for massive Pop~III clusters ($M_\star > 10^{4}\ \rm M_\odot$) given $N\ge 10$, where the cosmic average efficiency is $\sim 0.008$. 
We plan to conduct more detailed analyses on Pop~III SFE and its correlation with IMF in future work by applying our model to a cosmologically representative sample of Pop~III star-forming haloes. 

\subsection{Caveats and future improvements}
\label{sec:caveats}
The goal of this paper is to provide a flexible analytical framework that can predict the outcomes of Pop~III star formation from the properties of star-forming clouds and clarify the key physical processes that determine the final masses of Pop~III stars. As an initial version, our model is kept as simple as possible, while reproducing the key results of small-scale simulations of primordial star-forming clouds \citep{Hirano2014,Sugimura2023,Toyouchi2023}. It does not necessarily take into account all possible factors governing Pop~III star formation, and some model elements can be improved to better reflect the complexity in simulations and reality, as we discuss below.

\begin{enumerate}
        \item 
        The scaling relations used in our model (Sec.~\ref{sec:scale}) describe the (smoothed) median evolution for a large sample of Pop~III protostar systems focusing on the \textit{surviving} protostars. However, the evolution of individual systems shows strong stochasticity and diversity in simulations. For instance, the accretion rates of protostars fluctuate significantly at very short timescales for Pop~III star-forming clouds across a broad range of initial collapse rates $\dot{M}_{\rm in}\sim 10^{-4}-1\ \rm M_\odot\ yr^{-1}$, modulated by the interactions between protostars and the clumpy structure of the unstable star-forming disc  \citep{Sakurai2016,Hosokawa2016,Chon2020,Park2022,Park2024,Sugimura2020,Sugimura2023,Toyouchi2023,Prole2023bh,Reinoso2023}. This has interesting implications on the bloating mode of Pop~III star formation (Sec.~\ref{sec:mass}). For typical protostars in this regime with masses $\sim 1000-10^5\ \rm M_\odot$ and KH timescales $t_{\rm KH}\sim 100-1000\ \rm yr$, the bloating phase will terminate if the accretion rate stays below the critical rate $\dot{M}_{\star,\rm crit}\sim 0.01-0.04\ \rm M_\odot\ yr^{-1}$ for a duration longer than $\sim 10-100t_{\rm KH}\sim 10^{3}-10^{5}\ \rm yr$ \citep{Schleicher2013,Sakurai2015,Toyouchi2023,Reinoso2023} in which the star contracts to produce intense ionizing radiation. 
        Such contraction is more likely to occur at later stages as the accretion rate generally decreases with time. In the current model, we simply assume that the contraction occurs exactly when the (smoothed) median accretion rate drops below $\dot{M}_{\star,\rm crit}$, while in reality, the timing and the resulting final mass may have a large scatter due to the stochastic nature of this process. 
        We plan to consider such fluctuating accretion rates as well as protostar mergers and ejections in future work based on statistical analyses of 
        the protostar accretion/merger/ejection histories in simulations and/or physically motivated models for fragmentation via disc instability \citep[e.g.,][]{Tsukamoto2015,Inoue2020,Kimura2021}. In this way, we can explicitly follow the mass evolution of individual protostars and their circumstellar discs in a fully dynamical manner.
        
        \item We assume that Pop~III star formation happens via monolithic collapse with weak turbulence based on the simulations by \citet{Chon2021} for isolated primordial star-forming clouds with masses $M_{\rm c}\sim 2000\ \rm M_\odot$ \citep[see also][]{Higashi2022}. However, this assumption may not always hold in realistic environments of Pop~III star formation, especially for more massive clouds with violent collapse ($\dot{M}_{\rm in}\gtrsim 0.1\ \rm M_\odot\ yr^{-1}$). The presence of strong turbulence in star-forming clouds has two consequences: First, gas inflows onto the star-forming disc can be highly dynamical, deviating from the steady declining trend described by Eq.~\ref{mdot_m} \citep[see their fig.~6]{Regan2020}. Second, core fragmentation can break the cloud into multiple clumps each hosting their own star-forming discs and star clusters, which is particularly relevant when strong streaming motion between baryons and dark matter produces filamentary clouds \citep{Hirano2018sc,Hirano2023,Latif2022turb}. {These effects are expected to reduce the final masses of Pop~III stars formed. It is feasible to include turbulent fragmentation in our framework like what has been done for present-day star formation \citep[e.g.,][]{Larson1973,Hennebelle2008,Hopkins2012,Guszejnov2015,Thomasson2024}, but a larger sample of simulations is required to better understand the statistical properties of turbulent fragmentation under various conditions of primordial star formation}. 
        
        \item We adopt a simple power-law form with an invariant slope for the mass function of protostars, in the hope that it is flexible enough to qualitatively demonstrate the effects of multiplicity. Our model is not meant to reflect the reality, where the time evolution and shape of protostar mass function can be more complex, especially for SMS formation with high gas inflow rates ($\dot{M}_{\rm in}\gtrsim 0.1\ \rm M_\odot\ yr^{-1}$). Since high accretion rates \citep[$> \dot{M}_{\star,\rm crit}\sim 0.01-0.04\ \rm M_\odot\ yr^{-1}$,][]{Hosokawa2013,Haemmerle2018,Herrington2023,Nandal2023} cause significant expansion of some protostars during such violent collapse, super-competitive accretion and stellar collisions can cause a `class bifurcation' in the stellar population, where the total stellar mass is dominated by a small number of SMSs accompanied by a large number of low-mass stars \citep[e.g.,][]{Chon2020,Prole2023bh,Reinoso2023,Schleicher2023}. The exact form of the mass function of Pop~III (proto)stars (at the cloud scale) is still in debate, as different simulations produce highly divergent results which often suffer from poor statistics with a small number of realisations. We defer the consideration of more complex mass distributions to future work. 
        
        \item Related to the previous point (iii), the dependence of fragmentation outcomes on environmental factors (e.g., $v_{\rm bc}$, LW radiation, and halo dynamics) and cloud/disc properties (e.g., $\dot{M}_{\rm in}$, $M_{\rm c}$, $K$, $\gamma_{\rm eff}$, and $\eta$, or even the underlying temperature, density, and velocity profiles) regulated by larger (cosmic/halo)-scale physics is also ignored in our calculation of the $\hat{M}_{\star}$-$\dot{M}_{\rm in}$ relation. Recent simulations start to find hints for the existence of a non-trivial dependence \citep[see, e.g., fig.~4 in][]{Regan2024}. {For instance, it is shown by \citet[see their fig.~7]{Saavedra-Bastidas2024} that $f_2$ can be accurately predicted from the (cloud-scale) SFE $\epsilon_{\star}\equiv \hat{M}_{\star}/M_{\rm c}$ using a machine learning model trained on existing simulation results, which implies that $f_2$ strongly depends on $\dot{M}_{\rm in}$. }
        The parameters for cloud/disc properties may also correlate with each other, while they are treated as independent variables in our case. One needs to analyse a large set of (multi-scale) simulations systematically covering the diverse environments of Pop~III star formation to fully characterize the correlations between cloud/disc properties and multiplicity parameters and their dependence on the larger-scale condition of primordial star formation. Here, we simply estimate the range of possible effects of such correlations on the final masses of Pop~III star clusters with controlled numerical experiments (see Sec.~\ref{sec:multiplicity} and Appendix~\ref{apdx:details}). 
        
        \item Our model does not include the effects of magnetic fields, which have been intensively investigated in recent numerical studies \citep[e.g.,][]{McKee2020,Sharda2020,Sharda2021,Hirano2021,Prole2022,Saad2022,Stacy2022,Hirano2022,Higashi2024,Sadanari2023,Sadanari2024,Sharda2024}. Currently, (radiative) magneto-hydrodynamic simulations have converged on the picture that magnetic fields are significantly amplified by a turbulent dynamo and rotation to near equipartition with kinetic energy in the central region of primordial star-forming clouds, regardless of the initial field strength. Nevertheless, there is no census on the roles played by such strong magnetic fields during the star formation process, which may depend on the initial cloud properties. For instance, \citet[see their fig.~7]{Sharda2024} find that dynamically strong magnetic fields slow down accretion onto protostars in typical primordial star-forming clouds with $M_{\rm c}\sim 1000\ \rm M_\odot$. However, the simulations by \citet{Hirano2021} targeting more massive ($M_{\rm c}\sim 2\times 10^{6}\ \rm M_\odot$), atomic-cooling clouds show that efficient extraction of angular momentum by magnetic fields increases the accretion rate and enhances fragmentation and protostar mergers, facilitating the growth of the primary protostar. Besides, it is shown by several groups \citep{Sharda2020,Saad2022,Stacy2022,Hirano2022,Sadanari2024} that magnetic pressure and torques can efficiently suppress and even eliminate fragmentation, while \citet{Prole2022} find that the number and total mass of protostars are unaffected by saturated magnetic fields at equipartition. It is non-trivial to model magnetic fields explicitly in a simple analytical manner. However, their effects can possibly be absorbed into some existing parameters (e.g., $\dot{M}_{\rm in}$, $K$, $\gamma_{\rm eff}$, $\eta$, and $N$) if clearer trends can be deduced from simulations, which is an interesting topic for follow-up studies.
\end{enumerate}

\section{Summary and outlook}
We build an analytical model for Pop~III star formation to predict the final masses of Pop~III clusters and stars from the (initial) properties of star-forming clouds, using physically motivated prescriptions for the following physical processes and quantities: 
\begin{enumerate}
    \item[(1)] gas inflow onto the star-forming disc,
    \item[(2)] disc geometry and fragmentation, the resulting spatial, mass, and accretion rate distributions of protostars,
    \item[(3)] destruction of the star-forming disc/cloud by stellar feedback,
    \item[(4)] evolution of protostars regulated by their accretion histories. 
\end{enumerate}
The first two processes are expected to be closely related to the cloud-scale (initial) condition of star formation, while the last two processes lead to the termination of star formation (via depletion of gas supply or stellar collapse/explosion). Any self-consistent model of star formation must consider these four aspects. The basic idea of our model is to express them as functions of the total mass of protostars, so that the final mass can be easily obtained in a root-finding process.

To be specific, given the initial gas inflow rate $\dot{M}_{\rm in}$ onto the star-forming disc, we use power-law scaling relations between mass, size, and time to describe the (smoothed) median evolution of Pop~III protostar systems (Sec.~\ref{sec:scale}), which is combined with a phenomenological fragmentation model (see Sec.~\ref{sec:multiplicity} and Appendix~\ref{apdx:details}) to handle points (i) and (ii), based on the results of (magneto-) hydrodynamic simulations \citep{Liu2021wind,Sharda2020}.
For point (iii), we calculate the disc photo-evaporation rate as a function of the production rates of ionizing photons from individual protostars \citep{Tanaka2013} considering different spatial configurations of protostars and their circumstellar discs from point (ii). Besides, for point (iv), we include the bloating phase of protostar evolution with negligible ionizing power, triggered by high accretion rates \citep[$\gtrsim \dot{M}_{\star,\rm crit}\sim 0.01-0.04\ \rm M_\odot\ yr^{-1}$,][]{Hosokawa2013,Haemmerle2018,Herrington2023,Nandal2023}, and consider the finite lifetimes and maximum masses of stars before collapse/explosion (see Sec.~\ref{sec:mass}). 

Despite its simplicity, our model covers the full range of outcomes of Pop~III star formation known to date, from ordinary small ($M_\star\sim 10-2000\ \rm M_\odot$) clusters in molecular-cooling clouds to massive ($M_\star\gtrsim 10^{4}\ \rm M_\odot$) clusters containing supermassive ($m_\star\sim 10^{4}-3\times 10^{5}\ \rm M_\odot$) 
stars (SMSs) under violent collapse ($\dot{M}_{\rm in}\gtrsim 0.1\ \rm M_\odot\ yr^{-1}$) of atomic-cooling clouds, reproducing relevant simulation results \citep{Hirano2014,Sugimura2023,Toyouchi2023}. Moreover, our model specifically considers the interplay between feedback and fragmentation. We find that enhanced fragmentation tends to increase the final mass of the star cluster regulated by photo-ionization feedback under low gas inflow rates ($\dot{M}_{\rm in}\lesssim 0.1\ \rm M_\odot\ yr^{-1}$). However, it suppresses the formation of SMSs in higher-$\dot{M}_{\rm in}$ clouds not only by internal starvation \citep{Prole2022res} but also due to the decrease of final cluster mass caused by enhanced photo-ionization feedback from low-mass slowly-accreting ($\lesssim \dot{M}_{\star,\rm crit}$) stars. 

As an example, we apply our model to the Pop~III star-forming clouds in the main progenitors of typical haloes hosting high-$z$ luminous quasars with $M_{\rm h}\sim 10^{12}\ \rm M_\odot$ at $z\sim 6$ \citep{Li2021}, which predicts broad mass distributions for Pop~III clusters and stars extending to a few $10^5\ \rm M_\odot$ (Sec.~\ref{sec:mdis}). This shows that the formation of Pop~III massive clusters is common (among $\sim 20-70\%$ of the Pop~III star-forming haloes) in such biased ($\sim4\sigma$) regions, which favours the production of dynamical Pop~III BBH mergers with distinct gravitational wave signatures \citep{Wang2022,Liu2023sc,Mestichelli2024}. 
The corresponding SFE (ratio of the total stellar and gas masses in the halo) covers a broad range $f_\star\sim 0.001-0.1$, and is generally higher for more massive clusters. 
As the number of stars $N$ per cloud increases, the shape of the overall stellar mass distribution becomes more similar to that of the protostar mass distribution assumed in the fragmentation model (for individual clouds), while it closely follows the cluster mass distribution when fragmentation is inefficient ($N\lesssim 3$). 
This indicates that the overall IMF of Pop~III stars is regulated by both large-scale physics (embodied by the gas inflow rate distribution for Pop~III star-forming clouds) and small-scale fragmentation/feedback processes. 
When fragmentation is inefficient ($N\lesssim 3$), the resulting Pop~III SMSs as progenitors of heavy ($\gtrsim 10^{4}\ \rm M_\odot$) BH seeds can account for a significant fraction ($\gtrsim 7\%$) of observed luminous ($\gtrsim 10^{46}\ \rm erg\ s^{-1}$) 
quasars at $z\sim 6$ \citep[e.g.,][]{Greene2024,Matthee2024,Kokorev2023,Akins2024} powered by SMBHs with $m_{\rm BH}\gtrsim 10^8\ \rm M_\odot$. 
On the other hand, if we extend this trend to present-day star formation in a metal-enriched medium (by increasing $N$ and the slope $\alpha$ of the protostar mass distribution), where cooling and fragmentation are much more efficient compared with the Pop~III case, the overall IMF is mostly shaped by small-scale processes rather than large-scale physics, such that it becomes nearly universal as long as the IMF in individual star clusters is relatively insensitive to cloud/cluster properties\footnote{To what extent (or at which scale) the IMF of present-day stars remains universal is still in debate. Recent observations and simulations have discovered hints for a non-trivial dependence of the IMF on various properties of the star-forming environment \citep[e.g., metallicity, temperature, density, star formation rate, and cloud mass,][]{Marks2012,Jerabkova2018,Lacchin2020,Gunawardhana2011,Cheng2023,Dib2023,Grudic2023,Rusakov2023,Yan2023,Tanvir2024}.}.

Finally, we also discuss the potential missing pieces and directions to improve the model, such as stochastic evolution, core fragmentation, complex forms of protostar mass distribution, correlations between cloud/disc properties and multiplicity parameters, their dependence on the larger-scale condition of primordial star formation, and the effects of magnetic fields (Sec.~\ref{sec:caveats}). In conclusion, our cloud/disc-scale analytical model for Pop~III star formation can be 
incorporated into semi-analytical models \citep[e.g.,][]{Manrique2015,Griffen2018,Dayal2020,Visbal2020,Li2021,Lupi2021,Hartwig2022,Hegde2023,Nebrin2023,Bovill2024,Ventura2024,Feathers2024,Trinca2024} and cosmological hydrodynamic simulations \citep[e.g.,][]{Johnson2013,Smith2015,Skinner2020,Kulkarni2021,Schauer2021,Kiyuna2023,Yajima2022,Garcia2023,Lenoble2024,Sugimura2024} that focus on larger (cosmic/halo)-scale physics (e.g., structure formation, radiation backgrounds, baryons-dark matter streaming motion, and metal enrichment) to self-consistently follow the formation and feedback of Pop~III stars across all scales in a cosmologically representative volume. Such self-consistent modelling is necessary to make comprehensive predictions on the observational signatures and imprints of Pop~III stars and their BH remnants, which can be directly compared with multi-messenger observations of Cosmic Dawn to validate and advance our theories of early star/galaxy/BH formation.

\section*{Acknowledgements}
{The authors thank the anonymous referee for insightful comments and helpful suggestions. }
We thank Wenxiu Li\textsuperscript{\href{https://orcid.org/0000-0002-1044-4081}{\includegraphics[width=2.5mm]{orcid.png}}} for providing the data underlying \citet[]{Li2021}. We thank Piyush Sharda\textsuperscript{\href{https://orcid.org/0000-0003-3347-7094}{\includegraphics[width=2.5mm]{orcid.png}}} for providing the data for protostar accretion histories from the simulations shown in \citet{Sharda2020}. We thank Daisuke Toyouchi\textsuperscript{\href{https://orcid.org/0000-0003-3467-6079}{\includegraphics[width=2.5mm]{orcid.png}}} for useful discussions and Anastasia Fialkov for helpful comments on the manuscript. BL gratefully acknowledges the support of the Royal Society University Research Fellowship. Research at the Perimeter Institute for Theoretical Physics is supported in part by the Government of Canada through the Department of Innovation, Science and Economic Development Canada and by the Province of Ontario through the Ministry of Colleges and Universities. The development of this paper benefited from in-person discussions at the \href{https://events.perimeterinstitute.ca/event/58/}{Dark Matter, First Light} workshop hosted by the Perimeter Institute for Theoretical Physics and the \href{https://events.simonsfoundation.org/event/2102b8e2-a6a7-4bd2-b9b8-d41bb73c72b1/summary}{First Star VII} conference organized by the Flatiron Institute and Columbia University. We thank the organizers of these events for providing us platforms to exchange ideas. {This work excessively used the public packages \texttt{numpy} \citep{vanderWalt2011}, \texttt{matplotlib} \citep{Hunter2007}, and \texttt{scipy} \citep{2020SciPy-NMeth}. The authors wish to express their gratitude to the developers of these packages and to those who maintain them.}

\section*{Data Availability}
The code and data underlying this paper will be shared on reasonable request to the corresponding author.
 



\bibliographystyle{mnras}
\bibliography{ref} 




\appendix




\section{Detailed results in a larger parameter space}
\label{apdx:details}

\begin{figure}
    \includegraphics[width=1\columnwidth]{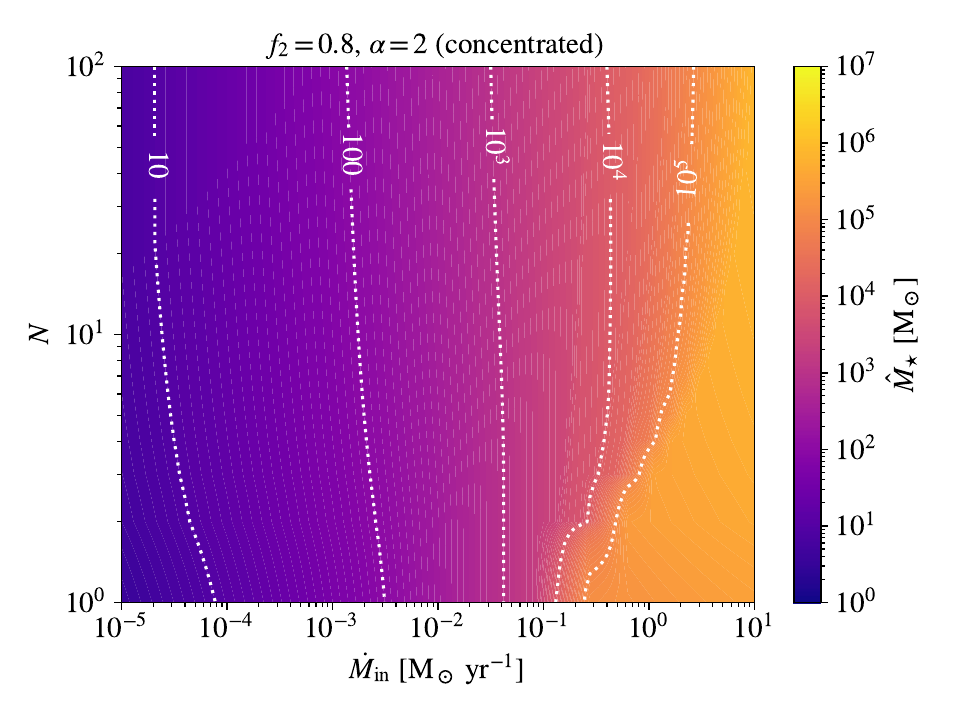}
    \vspace{-20pt}
    \caption{Map of $\hat{M}_{\star}$ in the $N$-$\dot{M}_{\rm in}$ space with other parameters fixed as in Fig.~\ref{fig:m_N}. {The dotted contours denote 5 specific levels of $\hat{M}_{\star}$ in units of $\rm M_\odot$.}}
    \label{fig:map_m_N}
\end{figure}

\begin{figure}
    \includegraphics[width=1\columnwidth]{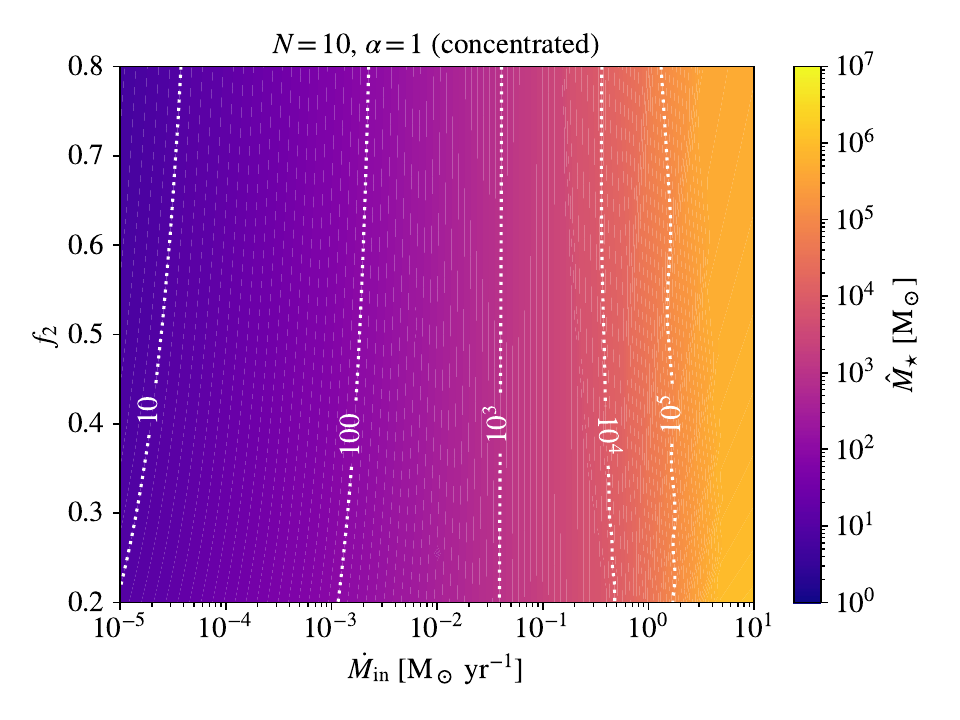}
    \vspace{-20pt}
    \caption{Map of $\hat{M}_{\star}$ in the $f_{2}$-$\dot{M}_{\rm in}$ space with other parameters fixed as in Fig.~\ref{fig:m_f2}. {The dotted contours denote 5 specific levels of $\hat{M}_{\star}$ in units of $\rm M_\odot$.}}
    \label{fig:map_m_f2}
\end{figure}

\begin{figure}
    \includegraphics[width=1\columnwidth]{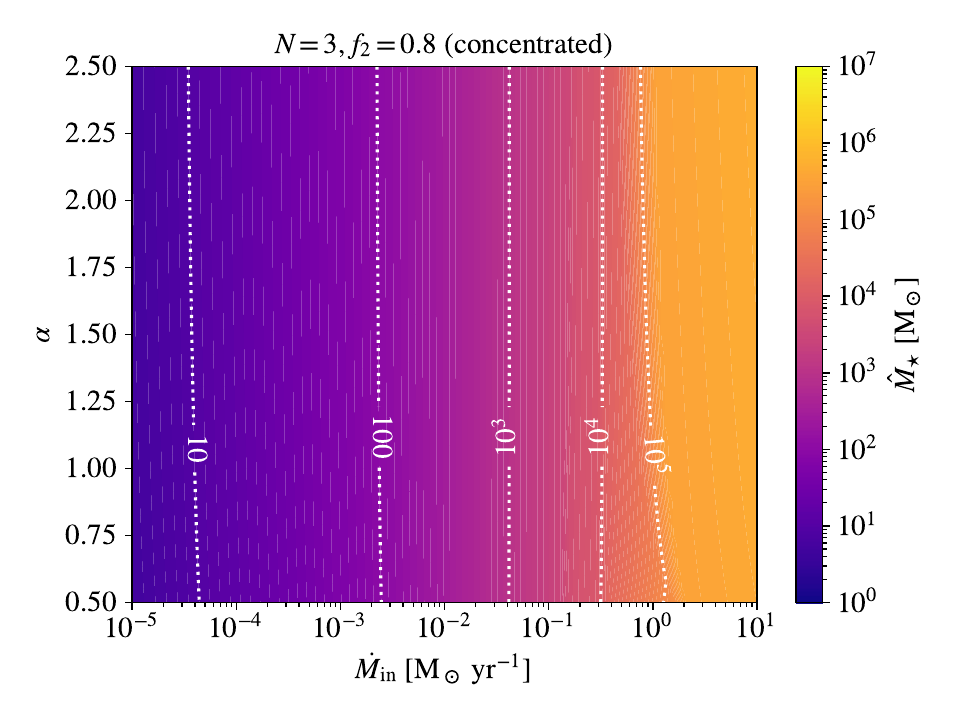}
    \vspace{-20pt}
    \caption{Map of $\hat{M}_{\star}$ in the $\alpha$-$\dot{M}_{\rm in}$ space with other parameters fixed as in Fig.~\ref{fig:m_alpha}. {The dotted contours denote 5 specific levels of $\hat{M}_{\star}$ in units of $\rm M_\odot$.}}
    \label{fig:map_m_alpha}
\end{figure}

In Sec.~\ref{sec:multiplicity}, we demonstrate the dependence of our results on mutiplicity parameters with select examples in the fiducial case where $N$ is independent of $M$ in the concentrated feedback scenario (Eq.~\ref{mdot_pe_whole}) given $\dot{M}_{\star,\rm crit}=0.04\ \rm M_\odot\ yr^{-1}$, 
$R_{\rm pe}\propto M^{1.25}$, and $\gamma_{\rm eff}=1.09$. 
In this Appendix, we first provide more detailed results in the fiducial case as maps of $\hat{M}_{\star}$ in Figs.~\ref{fig:map_m_N}-\ref{fig:map_m_alpha} covering the curves shown in Figs.~\ref{fig:m_N}-\ref{fig:m_alpha}. 
Then we vary the feedback and fragmentation prescriptions (Appendix~\ref{apdx:fdbk}) and the parameters governing the mass-size scaling relation and feedback strength (Appendix~\ref{apdx:scal}) to evaluate their impact on our results.

\subsection{Feedback and fragmentation prescriptions}
\label{apdx:fdbk}

In this section, we relax the assumptions of concentrated feedback (Eq.~\ref{mdot_pe_whole}) and constant $N$. As an extreme opposite to concentrated feedback, we consider the scenario where the distances between protostars are comparable to the disc size caused by outward migrations from accretion of angular momentum, as seen in recent radiative hydrodynamic simulations \citep[e.g.,][]{Sugimura2020,Sugimura2023,Park2022,Park2024}, so the ionizing flux from a protostar mostly affects its own circumstellar disc, and the total mass loss rate of photo-evaporation is the summation of the mass loss rates from the sub-discs around individual protostars: 
\begin{align}
    \dot{M}_{\rm pe}\simeq 0.015\ {\rm M_\odot\ yr^{-1}}\sum_{i}^{N}\biggl\{\left(\frac{\dot{q}_{{\rm ion},i}}{10^{52}\ \rm s^{-1}}\right)\left[\frac{R(m_{i})}{10^{4}\ \rm AU}\right]\biggl\}^{1/2}\ .\label{mdot_pe_subdisk}
\end{align}
Here, we assume that the sub-discs of individual protostars are always connected to each other (by spiral arms) so that the termination of protostar growth happens spontaneously when $\dot{M}_{\rm pe}$ exceeds $\dot{M}$. The maximum sub-disc radius $R_{\max,i}$ of each protostar $i$ is chosen to ensure that the total disc area does not exceed $\pi R_{\rm c}^{2}$:
\begin{align}
    R_{\max,i}=R_{\rm c}f_{i}^{\delta/\beta}/\left[\sum_{i}^{N}f_{i}^{2\delta/\beta}\right]^{1/2}\ ,
\end{align}
assuming that the size-mass scaling relation (Eq.~\ref{r_m_orig}) holds for each protostar\footnote{Alternatively, if all protostars have the same disc size, we simply use $R_{\max,i}=R_{\rm c}/N^{1/2}$.}. 
The photo-evaporation rate $\dot{M}_{\rm pe}$ is smaller in this scenario of distributed feedback due to the super-linear scaling relation between mass and size (Eq.~\ref{r_m}). 

Besides, it is shown by \citet{Susa2019} and \citet{Liu2021binary} that the (median) number of surviving protostars increases with time following a power-law scaling $N\propto t^{0.3}$ at least in the early stage\footnote{The number of fragments also increases with resolution without convergence up to densities of $10^{-6}\ \rm g\ cm^{-3}$ \citep{Prole2022res}.} in most simulations of Pop~III star-forming clouds. It is unclear whether this trend can hold till the end of star formation. In our default setup (Sec.~\ref{sec:multiplicity}), we assume that the number of surviving protostars has saturated at an input value \citep[e.g.,][]{Shima2021,Chon2021} independent of $M$ before the accretion of protostars is turned down by feedback \citep[i.e., $t_{\rm frag}<t_{\rm acc}$ in the formalism of][]{Liu2021binary}. 
Now, as an alternative model, we assume that $N$ keeps growing in time and introduce the dependence of $N$ on $M$ as
\begin{align}
    N=\max\left[1,N_{0}(M/M_{0})^{0.3/\beta}\right]\ ,\label{N_m}
\end{align}
using Eq.~\ref{m_t_final}, where $N_{0}$ is the new parameter defined as the number of protostars at $t=t_{0}$. We keep $f_{2}$ and $N_0$ fixed to the input value, and let $f_{1}(N,f_{2},\alpha)$ evolve with $M$ through the dependence on $N$, as fragmentation is expected to be important mainly for the smallest objects in the system. 

\begin{figure}
    \centering
    \includegraphics[width=1\columnwidth]{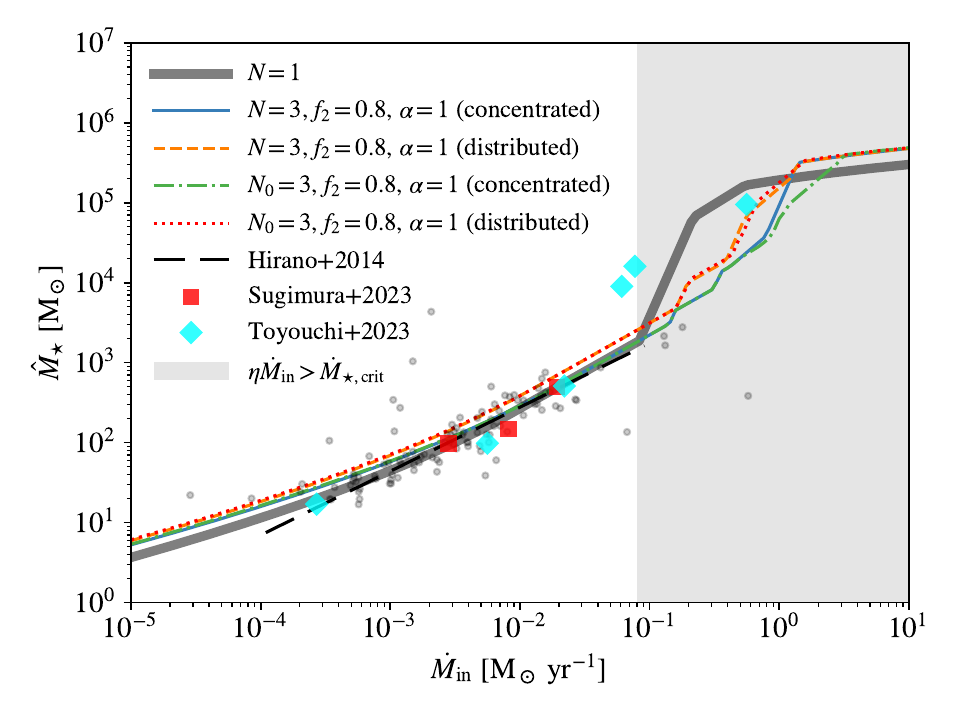}
    \vspace{-20pt}
    \caption{Same as Fig.~\ref{fig:m_N} but for four cases combining the treatments of $N$ and $\dot{M}_{\rm pe}$ given $N\ (N_{0})=3$, $f_{2}=0.8$, and $\alpha=1$. The solid (dash-dotted) and dashed (dotted) curves show the results without (with) time evolution of $N$ under concentrated and distributed feedback, respectively.}
    \label{fig:m_nevo}
\end{figure}

\begin{figure}
    \includegraphics[width=1\columnwidth]{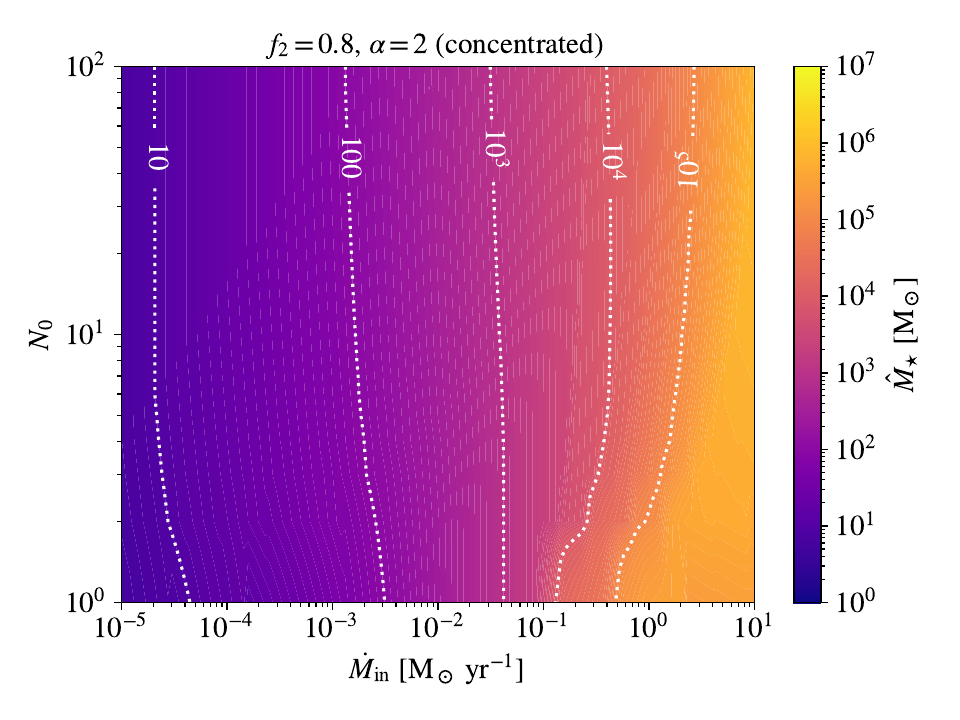}
    \vspace{-20pt}
    \caption{Map of $\hat{M}_{\star}$ in the $N_{0}$-$\dot{M}_{\rm in}$ space for the $N$ evolution model in Eq.~\ref{N_m}, with other parameters fixed as in Fig.~\ref{fig:m_N} (to be compared with Fig.~\ref{fig:map_m_N}). {The dotted contours denote 5 specific levels of $\hat{M}_{\star}$ in units of $\rm M_\odot$.}}
    \label{fig:map_m_N0}
\end{figure}

\begin{figure}
    \includegraphics[width=1\columnwidth]{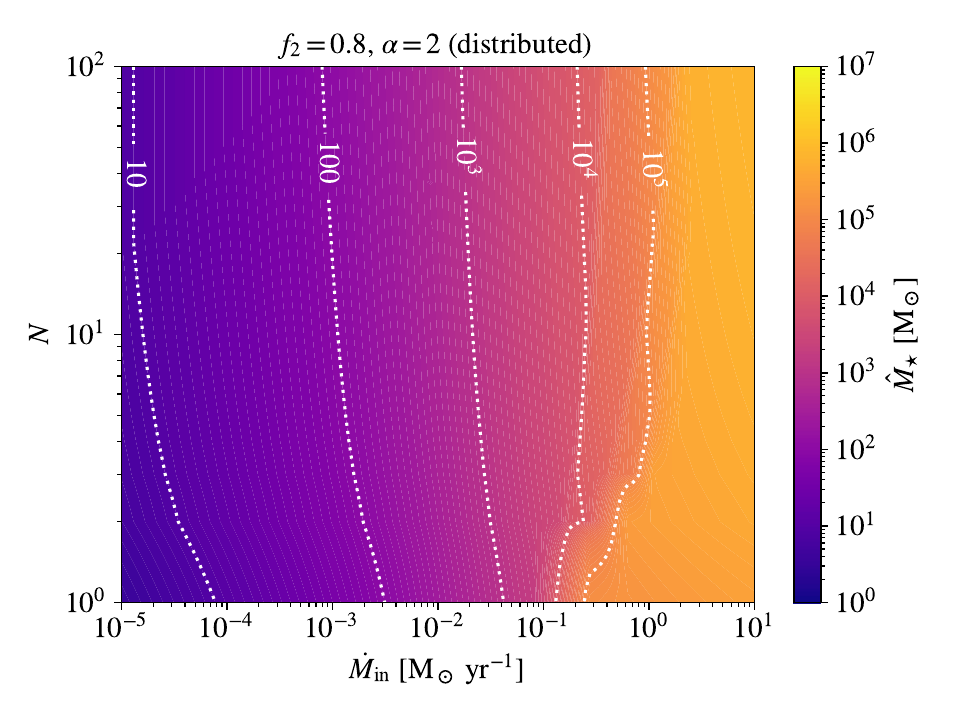}
    \vspace{-20pt}
    \caption{Map of $\hat{M}_{\star}$ in the $N$-$\dot{M}_{\rm in}$ under distributed feedback (Eq.~\ref{mdot_pe_subdisk}), with other parameters fixed as in Fig.~\ref{fig:m_N} (to be compared with Fig.~\ref{fig:map_m_N}). {The dotted contours denote 5 specific levels of $\hat{M}_{\star}$ in units of $\rm M_\odot$.}}
    \label{fig:map_m_N_dis}
\end{figure}

\begin{figure}
    \includegraphics[width=1\columnwidth]{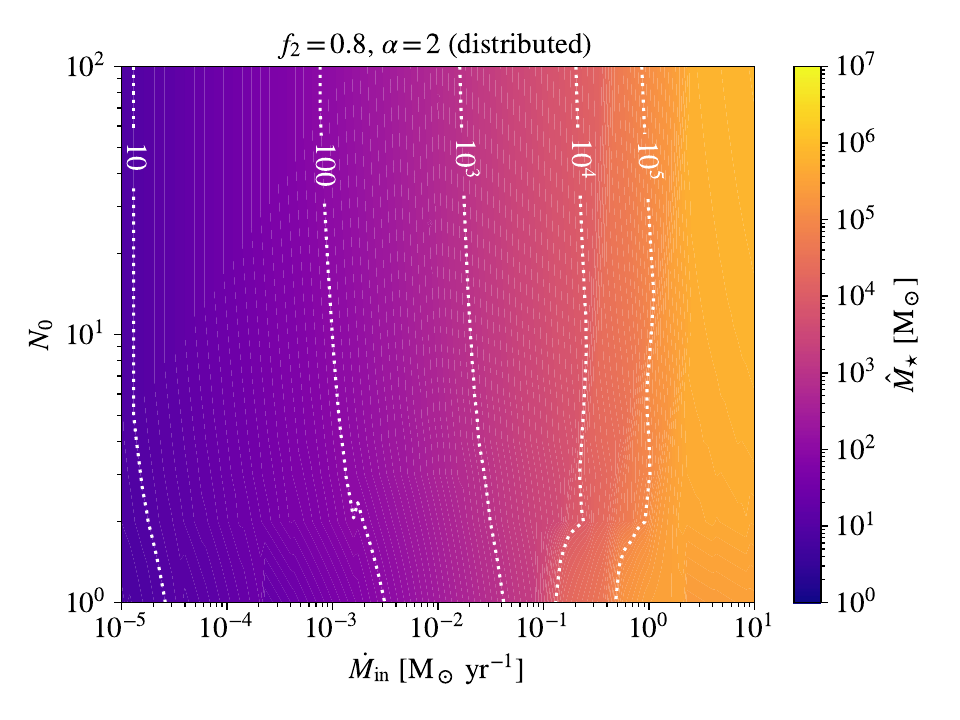}
    \vspace{-20pt}
    \caption{Map of $\hat{M}_{\star}$ in the $N_{0}$-$\dot{M}_{\rm in}$ space for the $N$ evolution model in Eq.~\ref{N_m} under distributed feedback (Eq.~\ref{mdot_pe_subdisk}), with other parameters fixed as in Fig.~\ref{fig:m_N} (to be compared with Fig.~\ref{fig:map_m_N}). {The dotted contours denote 5 specific levels of $\hat{M}_{\star}$ in units of $\rm M_\odot$.}}
    \label{fig:map_m_N0_dis}
\end{figure}

Fig.~\ref{fig:m_nevo} 
shows the effects of the treatments for $\dot{M}_{\rm pe}$ and $N$ for the exemplar multiplicity model with $N\ (N_0)=3$, $f_{2}=0.8$, and $\alpha=1$. As expected, $\hat{M}_{\star}$ 
increases when we consider distributed feedback, in particular for the intermediate regime, where the circumstellar discs of the most massive stars that undergo bloating are intact from photo-evaporation. On the other hand, allowing $N$ to increase with $M$ reduces $\hat{M}_{\star}$ 
in the intermediate regime under concentrated feedback. However, the effect is minor for distributed feedback, because when the disc sizes approach their upper limits ($M\gtrsim 2\times 10^{4}\ \rm M_\odot$), enhanced fragmentation increases the ionizing flux from small protostars that do not enter the bloating phase but meanwhile reduce the disc sizes, and the two effects cancel out each other in the disc evaporation rate. 

More detailed results with varying $N$ ($N_0$) in terms of maps of $\hat{M}_{\star}$ in the $N\ (N_{0})$-$\dot{M}_{\rm in}$ space for the three additional combinations of fragmentation and feedback prescriptions considered here are shown in Figs.~\ref{fig:map_m_N0}-\ref{fig:map_m_N0_dis}, to be compared with the fiducial case (Fig.~\ref{fig:map_m_N}).

\subsection{Mass-size scaling relation and feedback parameters}
\label{apdx:scal}

\begin{figure}
    \centering
    \includegraphics[width=1\columnwidth]{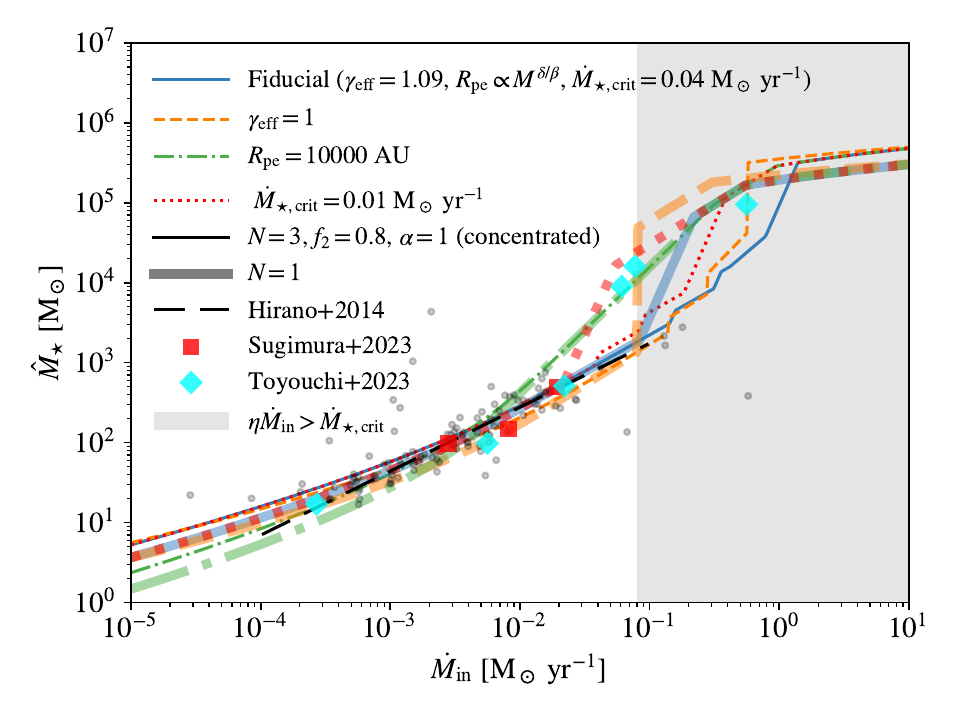}
    \vspace{-20pt}
    \caption{Same as Fig.~\ref{fig:m_N} but for four models with different parameters for the mass-size scaling relation and feedback strength. The solid curves show the results for the fiducial model with $\gamma_{\rm eff}=1.09$, $R_{\rm pe}\propto M^{\delta/\beta}$, and $\dot{M}_{\star,\rm crit}=0.04\ \rm M_\odot\ yr^{-1}$, while we adopt $\gamma_{\rm eff}=1$, $R_{\rm pe}=10^{4}\ \rm AU$, and $\dot{M}_{\star,\rm crit}=0.01\ \rm M_\odot\ yr^{-1}$ for the dashed, dash-dotted, and dotted curves, respectively, with the other parameters set to the fiducial values. In each case, the thin and solid curves show the results for the exemplar multiplicity model (with $N=3$, $f_{2}=0.8$, and $\alpha=1$), and the single-star model ($N=1$), respectively.}
    \label{fig:m_comp}
\end{figure}

Next, we explore the effects of the mass-size scaling relation and feedback parameters on our results by varying the polytropic index $\gamma_{\rm eff}$ of the effective EoS, the assumption on the disc photonevaporation scale $R_{\rm pe}$, and $\dot{M}_{\star,\rm crit}$ for the single-star case ($N=1$) and the exemplar multiplicity model with $N=3$ (constant), $f_{2}=0.8$, $\alpha=1$, and concentrated feedback, as shown in Fig.~\ref{fig:m_comp}. 
Here we consider three new models with $\gamma_{\rm eff}=1$, $R_{\rm pe}=10^{4}\ \rm AU$, and $\dot{M}_{\star,\rm crit}=0.01\ \rm M_\odot\ yr^{-1}$, varying one aspect each time with respect to the fiducial model ($\gamma_{\rm eff}=1.09$, $R_{\rm pe}\propto M^{\delta/\beta}$, and $\dot{M}_{\star,\rm crit}=0.04\ \rm M_\odot\ yr^{-1}$). The $\gamma_{\rm eff}=1$ model is meant to capture Pop~III star formation under extremely strong LW radiation \citep[with an intensity $J_{\rm LW}\gtrsim 1000$ in units of $10^{-21}\ \rm erg\ s^{-1}\ Hz^{-1}$,][]{Sugimura2014} or the cosmic microwave background at very high redshifts \citep[$z\gtrsim 500$,][]{Ito2024}, where $\rm H_2$ formation is significantly suppressed and the gas remains quasi-isothermal during collapse by atomic cooling \citep[see fig.~4 in][]{Li2021}.

\begin{figure}
    \includegraphics[width=1\columnwidth]{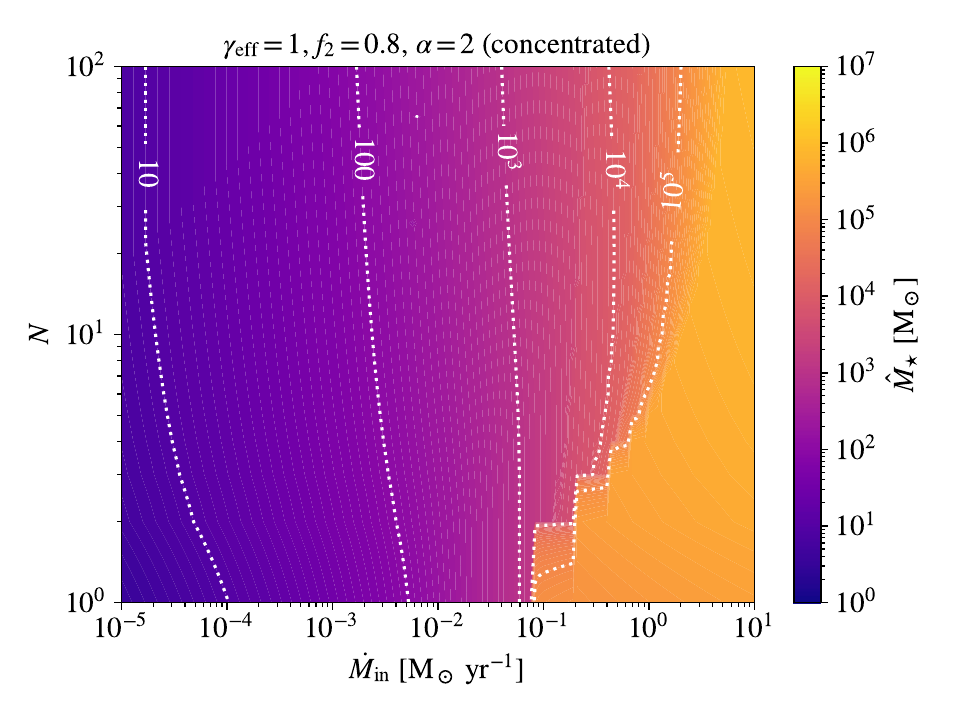}
    \vspace{-20pt}
    \caption{Map of $\hat{M}_{\star}$ in the $N$-$\dot{M}_{\rm in}$ space for $\gamma_{\rm eff}=1$ (instead of $\gamma_{\rm eff}=1.09$) with other parameters fixed as in Fig.~\ref{fig:m_N} (to be compared with Fig.~\ref{fig:map_m_N}). {The dotted contours denote 5 specific levels of $\hat{M}_{\star}$ in units of $\rm M_\odot$.}}
    \label{fig:map_m_N_gam1}
\end{figure}

\begin{figure}
    \includegraphics[width=1\columnwidth]{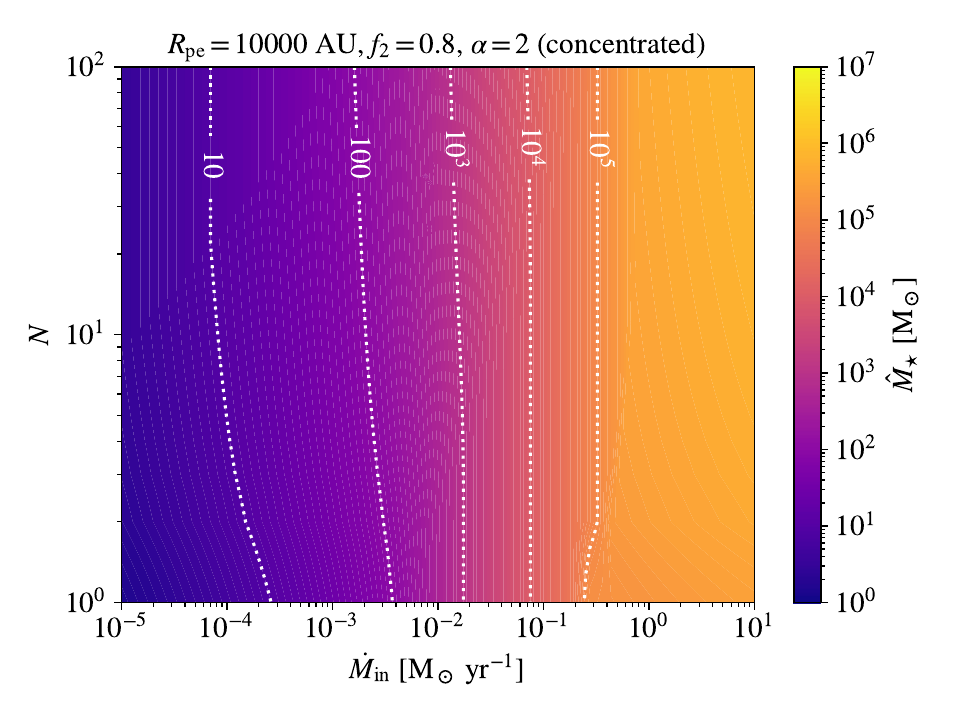}
    \vspace{-20pt}
    \caption{Map of $\hat{M}_{\star}$ in the $N$-$\dot{M}_{\rm in}$ space assuming a fixed disc photo-evaporation scale $R_{\rm pe}=10^{4}\ \rm AU$ (instead of the evolution model in Eq.~\ref{r_m}) with other parameters fixed as in Fig.~\ref{fig:m_N} (to be compared with Fig.~\ref{fig:map_m_N}). {The dotted contours denote 5 specific levels of $\hat{M}_{\star}$ in units of $\rm M_\odot$.}}
    \label{fig:map_m_N_rd10000}
\end{figure}

\begin{figure}
    \includegraphics[width=1\columnwidth]{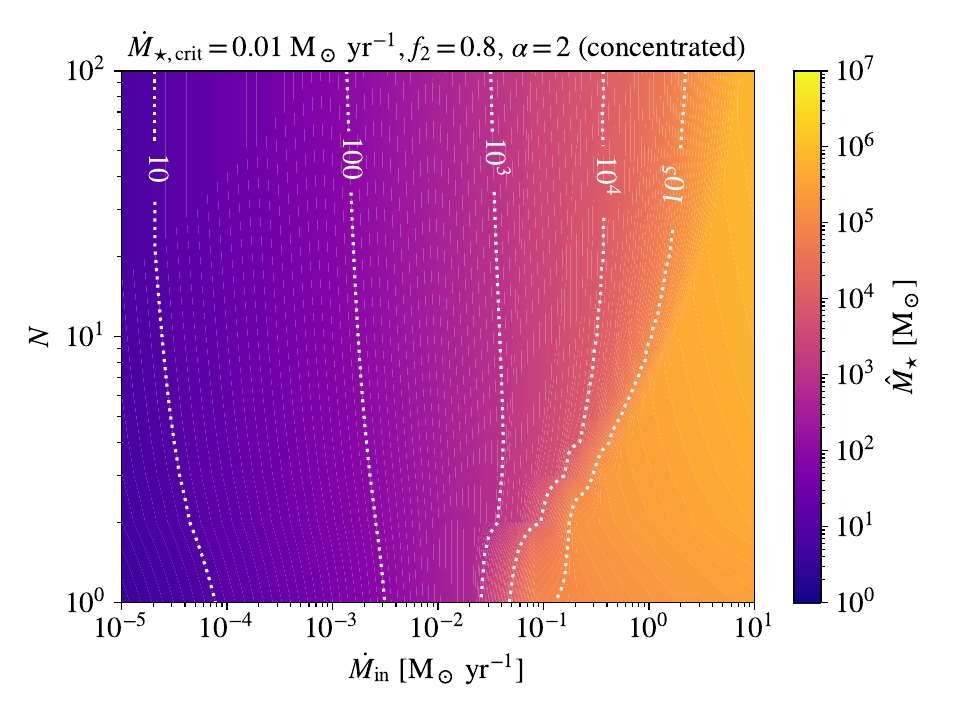}
    \vspace{-20pt}
    \caption{Map of $\hat{M}_{\star}$ in the $N$-$\dot{M}_{\rm in}$ space for $M_{\star,\rm crit}=0.01\ \rm M_\odot\ yr^{-1}$ (instead of $M_{\star,\rm crit}=0.04\ \rm M_\odot\ yr^{-1}$) with other parameters fixed as in Fig.~\ref{fig:m_N} (to be compared with Fig.~\ref{fig:map_m_N}). {The dotted contours denote 5 specific levels of $\hat{M}_{\star}$ in units of $\rm M_\odot$.}}
    \label{fig:map_m_N_mdotc0.01}
\end{figure}

If the star-forming gas is isothermal with $\gamma_{\rm eff}=1$, the accretion rate does not evolve with time and $M$, so in the single-star case, there is a sharp transition from the feedback-regulated regime to the lifetime/GRI-regulated regime (where the bloating phase lasts until the the star collapses) at $\eta\dot{M}_{\rm in}=\dot{M}_{\star,\rm crit}$. Therefore, $\hat{M}_{\star}$ is higher for $\gamma_{\rm eff}=1$ when the bloating phase is involved. However, $\hat{M}_{\star}$ becomes slightly smaller in the feedback-regulated regime because the disc size is smaller when $\beta$ increases (Eq.~\ref{r_m}) even though the accretion rate become higher (see Eq.~\ref{mdot_m}). The trends are similar when multiple protostars are considered. Now there is a jump in the $\hat{M}_{\star}$-$\dot{M}_{\rm in}$ curve each time the accretion rate of a protostar crosses $\dot{M}_{\star,\rm crit}$ such that it is in the bloating phase right before the end of star formation, which also occurs at a smaller $\dot{M}_{\rm in}$ compared with the $\gamma_{\rm eff}=1.09$ case. 

In the single-star case and the exemplar multiplicity model, using $R_{\rm pe}=10^{4}\ \rm AU$ instead of $R_{\rm pe}\propto M^{\delta/\beta}$ increases (reduces) $\hat{M}_{\star}$ for $\hat{M}_{\rm in}\gtrsim(\lesssim)0.006\ \rm M_\odot\ yr^{-1}$, while reducing $\dot{M}_{\star,\rm crit}$ facilitates the transition to the bloating or lifetime/GRI-regulated regime. 
With $R_{\rm pe}=10^{4}\ \rm AU$, the effects of multiplicity on $\hat{M}_{\star}$ vanish for $\hat{M}_{\rm in}\sim 0.01-0.2\ \rm M_\odot\ yr^{-1}$ in the feedback-regulated regime. These clouds are dominated by massive ($\gtrsim 200\ \rm M_\odot$) stars that follow the linear relation between $\dot{Q}_{\rm ion}$ and stellar mass, so the total production rate of ionizing photons and disc evaporation rate at a given total stellar mass $M$ hardly vary with $N$ at least for $N\le 100$. 
Interestingly, the models with $R_{\rm pe}=10^{4}\ \rm AU$ and $\dot{M}_{\star,\rm crit}=0.01\ \rm M_\odot\ yr^{-1}$ have similar results for $\dot{M}_{\rm in}\gtrsim 0.02\ \rm M_\odot\ yr^{-1}$ in the single-star case, consistent with the simulation results in \citet{Toyouchi2023}, where SMS formation starts at slightly lower $\dot{M}_{\rm in}$ compared with the fiducial model. This degeneracy is broken in the exemplar multiplicity model. 

To further illustrate the dependence on $N$, Figs.~\ref{fig:map_m_N_gam1}, \ref{fig:map_m_N_rd10000}, and \ref{fig:map_m_N_mdotc0.01} show the maps of $\hat{M}_{\star}$ in the $N$-$\dot{M}_{\rm in}$ space for the three new models with $\gamma_{\rm eff}=1$, $R_{\rm pe}=10^{4}\ \rm AU$, and $\dot{M}_{\star,\rm crit}=0.01\ \rm M_\odot\ yr^{-1}$, respectively, in comparison with the results of the fiducial model in Fig.~\ref{fig:map_m_N}. 


\section{Alternative parametrisation of multiplicity}
\label{apdx:param}

In Sec.~\ref{sec:multiplicity}, we build a phenomenological multiplicity model with four variables $\alpha$, $N$, $f_{1}$, and $f_{2}$, which are correlated with each other via Eq.~\ref{f1}, so only three of them can be chosen as free model parameters. 
Here, $\alpha$ and $N$ have clear physical meanings as the slope of mass distribution and target number of stars and therefore, they should be adopted as model parameters. $f_{1}$ and $f_{2}$ are related to the least and most massive star in the cluster, respectively. As we are mainly concerned with the formation of SMSs as the most massive objects in the clusters, we use $f_{2}$ as the last free parameter by default. 

Motivated by possible constraints on the minimum mass of metal-free stars by stellar archaeology \citep[e.g.,][]{Salvadori2007,Frebel2015,Hartwig2015,Komiya2016,Ishiyama2016,Magg2018,Magg2019,Dutta2020,Rossi2021}, here, we consider an alternative parametrisation of multiplicity by adopting $m_{\min}$ as the third free parameter (in replacement of $f_{2}$) with $f_{1}=m_{\min}/M$ given the total mass $M$. Once $f_{1}$ is known according to $m_{\min}$, $f_{2}=f_{2}(N,f_1,\alpha)$ is derived using Eq.~\ref{f1}. Since the solution of $f_{2}$ may not exist for the input (target) value of $N$, we constrain $N$ with an upper limit $N_{\max}=M/m_{\min}$ and a lower limit $N_{\min}$ corresponding to $f_{2}(N_{\min},f_1,\alpha)=1$. Below, we briefly explore the effects of multiplicity with the new parametrisation under the default setup (constant $N$, concentrated feedback, $\gamma_{\rm eff}=1.09$, $R_{\rm pe}\propto M^{1.25}$, $\dot{M}_{\star,\rm crit}=0.04\ \rm M_\odot\ yr^{-1}$, and $\eta=0.5$).

\begin{figure}
    \centering
    \includegraphics[width=1\columnwidth]{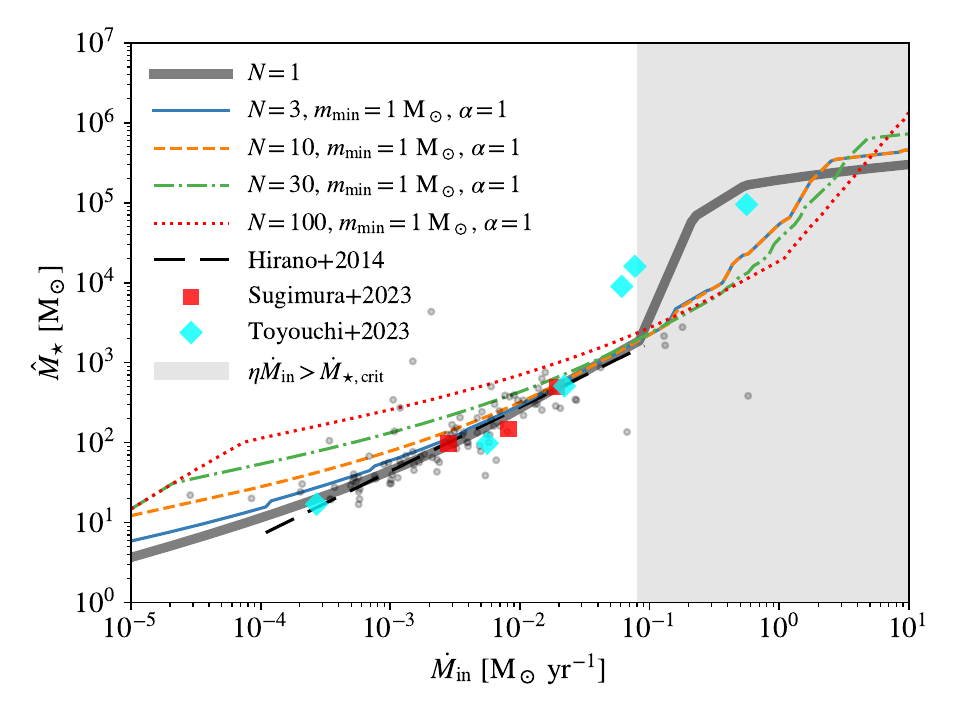}
    \vspace{-20pt}
    \caption{Final total stellar mass as a function of initial gas inflow rate 
    for $N=3$ (thin solid), 10 (dashed), 30 (dash-dotted), and 100 (dotted) with $m_{\min}=1\ \rm M_\odot$ and $\alpha=1$. The results for the single-star model ($N=1$) is shown with the thick solid curve for comparison. The fitting formula of final stellar mass (Eq.~\ref{h14}) based on 2D (single-star) simulations from \citet[see their fig.~14]{Hirano2014} is shown with the long-dashed line, and the dots show the underlying data. The results of the 3D AMR simulations in \citet{Sugimura2023} are denoted by the squares, while those of the 3D spherical simulations in \citet{Toyouchi2023} are shown with the diamonds.}
    \label{fig:m_N_mmin}
\end{figure}

\begin{figure}
    \centering
    \includegraphics[width=1\columnwidth]{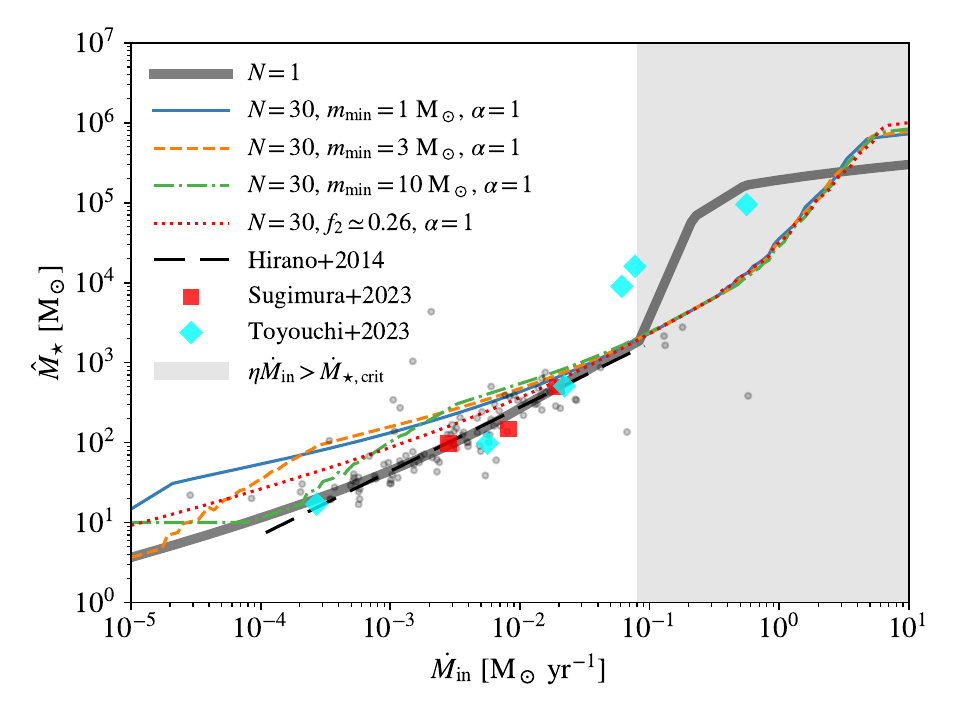}
    \vspace{-20pt}
    \caption{
    Same as Fig.~\ref{fig:m_N_mmin} but for $m_{\min}=1$ (thin solid), 3 (dashed), and $10\ \rm M_\odot$ (dash-dotted), with $N=30$ and $\alpha=1$. We also show the results for the default parametrisation with a fixed $f_{2}\sim 0.26$ as the dotted curve, which agree well with the results for the new parametrisation at $\dot{M}_{\rm in}\gtrsim 0.1\ \rm M_\odot\ yr^{-1}$ where $\hat{M}_{\star}\gtrsim 2000\ \rm M_\odot$ and $f_{1}$ is close to zero.}
    \label{fig:m_mmin}
\end{figure}

\begin{figure}
    \centering
    \includegraphics[width=1\columnwidth]{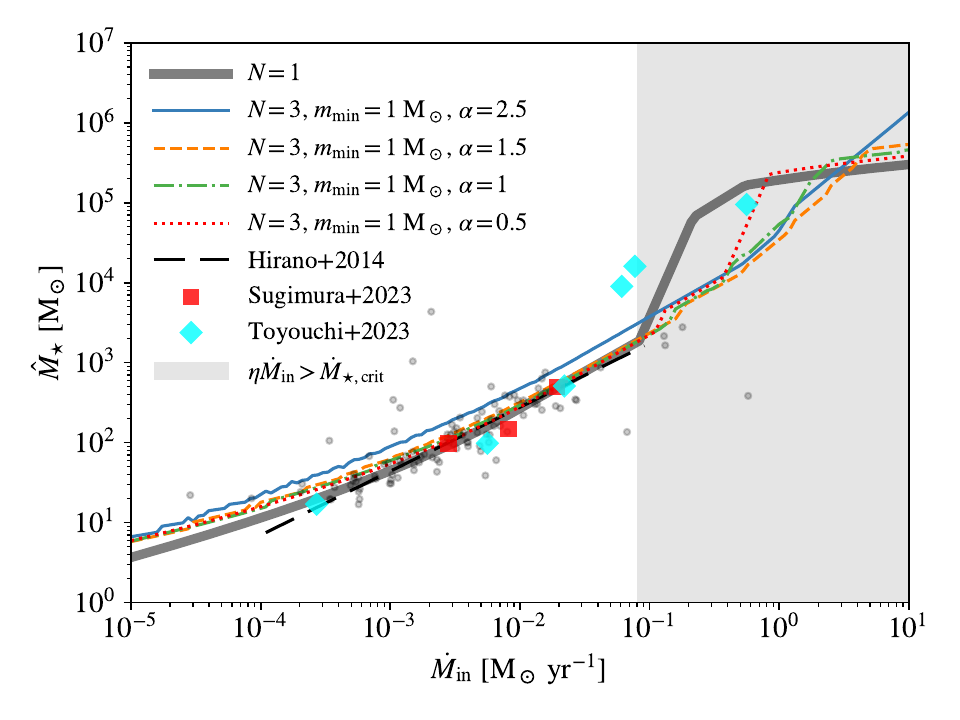}
    \vspace{-20pt}
    \caption{
    Same as Fig.~\ref{fig:m_N_mmin} but for $\alpha=2.5$ (thin solid), 1.5 (dashed), 1 (dash-dotted), and 0.5 (dotted) with $N=3$ and $m_{\min}=1\ \rm M_\odot$.}
    \label{fig:m_alpha_mmin}
\end{figure}

Fig.~\ref{fig:m_N_mmin} shows the evolution of $\hat{M}_{\star}$ with $\dot{M}_{\rm in}$ for $N=3$, 10, 30, and 100 given $\alpha=1$ and $m_{\min}=1\ \rm M_\odot$. We adopt $m_{\min}=1\ \rm M_\odot$ as the fiducial choice based on observations \citep{Salvadori2007,Hartwig2015,Magg2019,Rossi2021}. 
The trends are similar to the case where $f_{2}$ is fixed (Fig.~\ref{fig:m_N}), as $\hat{M}_{\star}$ becomes higher in the feedback-regulated regime ($\dot{M}_{\rm in}\lesssim 0.1\ \rm M_\odot\ yr^{-1}$) when $N$ increases, showing a slower transition to the lifetime/GRI-regulated regime. However, the effects of varying $N$ are stronger with the new parametrisation for $N\gtrsim 10$ because $f_{2}$ rapidly decreases with $N$ given a fixed $m_{\min}$, which reduces the ionization flux at $\dot{M}_{\rm in}\lesssim 0.1\ \rm M_\odot\ yr^{-1}$ and the mass fraction of stars in the bloating phase at $\dot{M}_{\rm in}\gtrsim 0.1\ \rm M_\odot\ yr^{-1}$. The final mass is even regulated by the gas supply for $N\gtrsim 10$ at the low $\dot{M}_{\rm in}$ end. 
Interestingly, the results for $N=3$ and 10 are identical at $\dot{M}_{\rm in}\gtrsim 0.1\ \rm M_\odot\ yr^{-1}$ because $N_{\min}\simeq 10$ in this regime ($M\gtrsim 2000\ \rm M_\odot$). 

The dependence on $m_{\min}$ is illustrated in 
in Fig.~\ref{fig:m_mmin} for $m_{\min}=1$, 3, and 10 $\rm M_\odot$ given $N=30$ and $\alpha=1$. Here the evolution of $\hat{M}_{\star}$ with $\dot{M}_{\rm in}$ has two stages. 
When $\dot{M}_{\rm in}$ is low, $N_{\max}$ is below the input target value of $N$ and increases with $\dot{M}_{\rm in}$, which explains the rapid increase of $\hat{M}_{\star}$ with $\dot{M}_{\rm in}$. For $m_{\min}=10\ \rm M_\odot$, $\hat{M}_{\star}=m_{\min}$ is achieved at $\dot{M}_{\rm in}\lesssim 10^{-4}\ \rm M_\odot$ where $N_{\max}\sim 1$. Once $N_{\max}$ excesses the target value of $N$ with higher $\dot{M}_{\rm in}$, the increase of $\hat{M}_{\star}$ first slows down (within the no-bloating feedback-regulated regime) before accelerating again when the bloating effect kicks in. In this stage, the dependence on $m_{\min}$ is minor since $f_{1}$ is close to 0 for all the cases considered here, whose results are well reproduced by the model with a fixed $f_{2}\simeq 0.26$ from the default parametrisation, given $\lim_{f_{1}\rightarrow 0}\left[f_{2}(N=30,f_{1},\alpha=1)\right]\simeq 0.26$. 

Finally, we consider the dependence on $\alpha$, as shown in Fig.~\ref{fig:m_alpha_mmin} with $N=3$ and $m_{\min}=1\ \rm M_\odot$ for $\alpha\sim 0.5-2.5$. 
At $\dot{M}_{\rm in}\lesssim 0.1\ \rm M_\odot\ yr^{-1}$, $\hat{M}_{\star}$ is higher when the mass distribution is more bottom-heavy, showing a consistent but stronger trend compared with the default case with fixed $f_{2}$ (Fig.~\ref{fig:m_alpha}), which reflects the weaker ionizing power of smaller stars. The trend will be even stronger if $N$ is larger. 
However, for $\dot{M}_{\rm in}\gtrsim 0.1\ \rm M_\odot\ yr^{-1}$, increasing $\alpha$ with a fixed $m_{\min}$ has a similar effect as increasing $N$ while fixing $f_{2}$ (and $\alpha$, see Fig.~\ref{fig:m_N}) for $\alpha\sim 0.5-2$. The reason is that the input value $N=3$ is no longer effective here as $N_{\min}> 3$ and $N_{\min}$ increases with $\alpha$. In the extreme case of $\alpha=2.5$ strongly dominated by low-mass stars, the final mass is always regulated by photo-evaporation feedback for $\dot{M}_{\rm in}\le 10\ \rm M_\odot\ yr^{-1}$, which is weaker when $\alpha$ is larger, so $\hat{M}_{\star}$ is higher than that for $\alpha=1.5$, holding the trend at $\dot{M}_{\rm in}\lesssim 0.1\ \rm M_\odot\ yr^{-1}$. 

\bsp	
\label{lastpage}
\end{document}